\documentclass[12pt]{article}%%%Put [range] option here for references citations to
\usepackage{amsmath}
\usepackage{amssymb}
\usepackage{a4wide}
\usepackage{graphicx}

\begin{document}

% \input epsf.tex    %<-If you need EPS figures to be
%                    %  called in {figure} environment for PC
% \input epsf.def   %<-If you need EPS figures to be
%                    %  called in {figure} environment for Macintosh
% 
% \input psfig.sty

\newcommand{\affiliation}[1]{#1\\[12pt]}
\date{}

% \jname{..}
% \jyear{2000}
% \jvol{}
% \ARinfo{1056-8700/97/0610-00}

\title{\sc Hard Processes in Proton-Proton Collisions at the Large Hadron
  Collider\footnote{Invited contribution to the Annual Review of
    Nuclear and Particle Science.}}

% \markboth{Butterworth, Dissertori \& Salam}{Hard Processes in pp
%   Collisions at the LHC}

\author{Jonathan M. Butterworth\\
\affiliation{\it
Department of Physics and Astronomy, \\\it
University College London, Gower St., London, WC1E 6BT, UK
}
G\"unther Dissertori\\
\affiliation{\it
Institute for Particle Physics, ETH Zurich, 8093 Zurich, Switzerland
}
Gavin P. Salam\\
\affiliation{\it
  CERN, Physics Department, Theory Unit, 1211 Geneva 23, Switzerland\\\it
  Department of Physics, Princeton University, Princeton, NJ 08544, USA\\\it
  LPTHE, UPMC and CNRS UMR 7589, 75252 Paris cedex 05, France
}
}

% \begin{keywords}
%   LHC, Hard Processes, Hadron Collisions, Jets, Vector Bosons, Heavy Quarks
% \end{keywords}

\maketitle

\begin{abstract}
  The measurement of hard scattering processes, meaning those with
  energy scales of more than a few GeV, is the main method by which
  physics is being explored and extended by the experiments at the
  Large Hadron Collider.  We review the principal measurements made so
  far, and what they have told us about physics at the energy
  frontier.
\end{abstract}

\section{Introduction}

The Large Hadron Collider (LHC) at CERN has been colliding protons
together at the unprecedented centre-of-mass energy of 7 TeV since the
$30^{\rm th}$ of March 2010. 
The main goal of the LHC is to explore physics above the electroweak
symmetry breaking (EWSB) scale.  This scale, which is roughly between
100 and 1000~GeV, is special for several related reasons.  The
mechanism for EWSB in the Standard Model (SM) is related to the Higgs
field, which has a vacuum expectation value of 246~GeV. If this is
realized in nature, the Higgs boson mass must lie in the range
accessible at the LHC. If not, other physics must enter to break the
symmetry and generate the masses of fundamental particles.  The $W$ and
$Z$ bosons and the top quark all have masses in this range --- in the
case of the bosons at least this is directly related to the symmetry
breaking itself --- and are thus copiously produced at the LHC.  Other
new physics, postulated to address perceived weaknesses in the
Standard Model, may also enter the picture. In general, the LHC is
exploring not only a new energy range, the highest ever studied, but
also a qualitatively new region of physics, where the electroweak
symmetry is restored.  For all these reasons, the detailed, precise
study of hard scattering processes at the LHC --- those involving a
momentum transfer that is large compared to the proton mass --- is a
critically important task, so as to establish whether our
understanding of fundamental physics remains valid in this new region,
or requires extension.

In 2010, the LHC delivered an integrated luminosity of nearly
50~pb$^{-1}$ to each of the two ``general purpose'' experiments, ATLAS
\cite{Aad:2008zzm} and CMS \cite{:2008zzk},
the equivalent of about 5 trillion proton-proton collisions each.
The LHC also provided data to the more specialised
LHCb \cite{Alves:2008zz} and ALICE \cite{Aamodt:2008zz} experiments,
including a period of heavy ion (lead-lead) collisions. 
In 2011, LHC luminosity increased rapidly such that, by the end of the
year, ATLAS and CMS had collected almost a factor 100 more data.
Ultimately, it is foreseen that the LHC will collect several hundred, or
even thousand, fb$^{-1}$ at a centre-of-mass energy of 14~TeV.

Given that this review focuses on hard scattering processes,
measurements from ATLAS and CMS naturally dominate, insofar as LHCb is
primarily targeted at (bottom) flavour physics, and ALICE at heavy-ion
physics.
To understand the capabilities of the LHC experiments it is useful to
be aware of the main kinematic variables used to characterize
particle momenta at hadron collisions. ``Hardness'' is usually specified in
terms of the momentum component that is transverse to the beam, $p_T$.
Direction tends to be expressed in terms of the pseudorapidity,
defined as $\eta = -\ln \tan \frac{\theta}{2}$ where $\theta$ is a
particle's polar angle w.r.t.\ the beam, together with the azimuthal angle
$\phi$.\footnote{This particular choice is motivated by the fact that
  pseudorapidity differences between two massless particles are
  invariant under longitudinal Lorentz boosts. The related variable
  rapidity, $y=\frac12 \ln\frac{E+p_z}{E-p_z}$, is often used instead and
  extends the boost invariance property to particles of arbitrary
  mass.}
The ATLAS and CMS detectors provide the most detailed measurements in
their central regions, $|\eta|\lesssim 2.5$, where specific detector
components provide highly reliable identification and measurement of
electrons, muons, photons and jets, as well as tagging of $\tau$s and
$B$-hadrons.
Calorimetry allows measurement of jets (and electrons and photons with
more limited or no identification capability) up to $|\eta|\lesssim 5.0$.
This broad coverage also facilitates determinations of missing
transverse energy (MET).
Typical lepton and photon transverse momentum thresholds applied
in measurements of hard processes are around $20$~GeV. Similarly,
thresholds of $\sim30$~GeV define the lower phase space boundary for
jet reconstruction.  If
information from the tracking systems is combined with calorimeter
signals for jet finding, such thresholds can be lowered to
$\sim15$~GeV without a significant degrading of performance.
Note also that the LHCb experiment, though not primarily targeted towards hard
processes, provides unique measurement possibilities for leptons and
$B$-hadrons at high rapidities, $2 < \eta < 4.5$.

\begin{figure}[htbp]
\begin{center}
  \includegraphics[height=14pc]{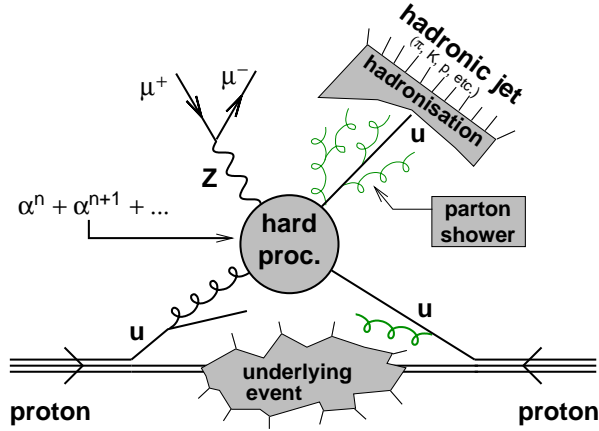}
\end{center}
\caption{Schematic representation of a proton-proton collision, 
  involving a quark-gluon scattering that leads to a final state
  consisting of a $Z$ boson and a hard jet.}
\label{fig:hard-scatter}
\end{figure}

Experimental results on hard processes are typically compared to
predictions of perturbative Quantum Chromodynamics (QCD).
Hard scattering processes probe distance scales far below the radius
of the proton, and thus are best understood as collisions between the
constituent quarks and gluons (generically called partons) of the
proton, as depicted in Fig.~\ref{fig:hard-scatter}. 
A typical calculation of a cross section thus consists of a term that
describes the partonic scatter (with cross section $\hat\sigma$) and
factors for the incoming flux of partons (the parton density functions
(pdfs) $f_{i,p}$), as cast in the general expression
\begin{equation}
  \label{eq:general-cross-section}
  \sigma(pp\rightarrow X) = \sum_{i,j} \int dx_1 dx_2\, 
     f_{i,p}(x_1,\mu_F^2)\, f_{j,p}(x_2,\mu_F^2)\, 
     \hat\sigma_{ij\rightarrow X}(x_1 x_2 s, \mu_R^2, \mu_F^2) \; .
\end{equation}
Here the sum runs over all possible initial-state partons, with
longitudinal momentum fractions $x_{1,2}$, that can give rise to a
final state $X$ at a centre-of-mass energy of $\sqrt{x_1 x_2 s}$.
Furthermore, the renormalization (factorization) scales $\mu_R^2\,
(\mu_F^2)$ appear in the expressions if they are obtained from
truncated expansions in the strong coupling constant.  
Under the assumption of factorization, which is proven for some
processes, the parton densities are universal at a given resolution,
or momentum scale. In addition, their evolution with scale is
determined by the strong interaction (QCD) and for hard enough scales
it can be calculated using perturbative techniques.  Thus, over a wide
range of momentum fractions and scales, the parton densities are
rather well known, in particular due to precise measurements of deep
inelastic lepton-proton scattering (see for example
\cite{Alekhin:2011sk} and references therein). The parametrizations
of pdfs based on fits incorporating the broadest range of data are
those from the CTEQ~\cite{Lai:2010vv}, MSTW~\cite{Martin:2009iq} and
NNPDF~\cite{Ball:2011uy} collaborations.

The hard cross section~$\hat \sigma$ may be calculated at leading
order (LO) in the strong coupling $\alpha_s$, or incorporating
next-to-leading (NLO) or even next-to-NLO (NNLO) corrections.
Such ``fixed-order'' predictions are associated with final states with
a small number of hard partons.
Physically however, shortly after being produced, hard partons
repeatedly radiate low-energy and collinear gluons, a process known as
a parton shower.
Calculations of parton showers, together with models for the partons'
transition to hadrons (hadronization) form the basis of widely used
general-purpose Monte Carlo simulation programs such as
Pythia~\cite{Sjostrand:2006za}, Herwig~\cite{Bahr:2008pv} and
Sherpa~\cite{Gleisberg:2008ta}, which give realistic descriptions of
events with all final-state particles.
It is common for parton showers to be matched with multi-leg
tree-level matrix elements~\cite{Alwall:2007fs} (e.g.\ from
Alpgen~\cite{Mangano:2002ea} or MadGraph~\cite{Alwall:2011uj}, or
directly incorporated in Sherpa) and also with NLO matrix elements, in
particular using methods known as MC@NLO~\cite{Frixione:2002ik} and
POWHEG~\cite{Nason:2004rx}.
As we will see, the LHC results are usually compared either to NLO or
NNLO fixed-order predictions or to results from parton-shower
programs, with or without matching.

A comprehensive introduction to the theoretical description of hard
scattering processes at hadron colliders can be found, eg., in Refs.\
\cite{Dissertori:2003pj,Ellis:1991qj,Nakamura:2010zzi,Buckley:2011ms}.

\section{Jets}

By far the most common hard process in hadron collisions is
the scattering of partons off each other. 
This is a consequence both of the high density of gluons in the proton
and the fact that the QCD coupling is significantly larger than the
electroweak couplings.
Still, at hard scales the QCD coupling's value is sufficiently small
that perturbative techniques can be used (the expansion in powers
of $\alpha_s$ is generally stable).  

A high energy quark or gluon emitted from such a high energy
scatter will not in the end appear in the detector, since as it
reaches larger distances from the rest of the proton, the QCD
force becomes stronger.  
Successively lower-energy (softer) gluons may be radiated, often at
small angles relative to the original parton, until a point where a
non-perturbative transition causes the partons bind into
colour-neutral hadrons.
The result is a more-or-less collimated
``jet'' of hadrons whose collective energy and momentum reflect at
some level those of the initial scattered parton. The hadrons can be
combined using various ``jet algorithms'' to allow this correspondence to be
made reproducibly and with a degree of precision. The LHC experiments use
the anti-$k_t$ algorithm~\cite{Cacciari:2008gp}, which is collinear
and infrared safe, meaning that the resulting hard jets are not
substantially affected by the small-angle (collinear) and soft splittings
that occur in a parton shower.
This characteristic is important also because it ensures that one
obtains finite results at every order in perturbation theory.

The reconstructed jet momenta are inputs for
measurements of, eg., jet $p_T$ distributions. Typically these are
steeply falling functions, therefore very sensitive to the precise
knowledge of the absolute momentum/energy scale and resolution. By
now, the jet energy scale uncertainties are controlled at the $1-3\%$
level \cite{Chatrchyan:2011ds,:2011he}, depending on jet momentum and
rapidity, and constitute the dominant systematic error in most jet
cross section measurements. 

In summary: jet cross sections give the
first opportunity to confront SM calculations with data at the highest
energies.

\subsection{Jet Cross Sections}

\begin{figure}[htbp]
\begin{center}
\begin{tabular}{lr}
\includegraphics[height=17pc]{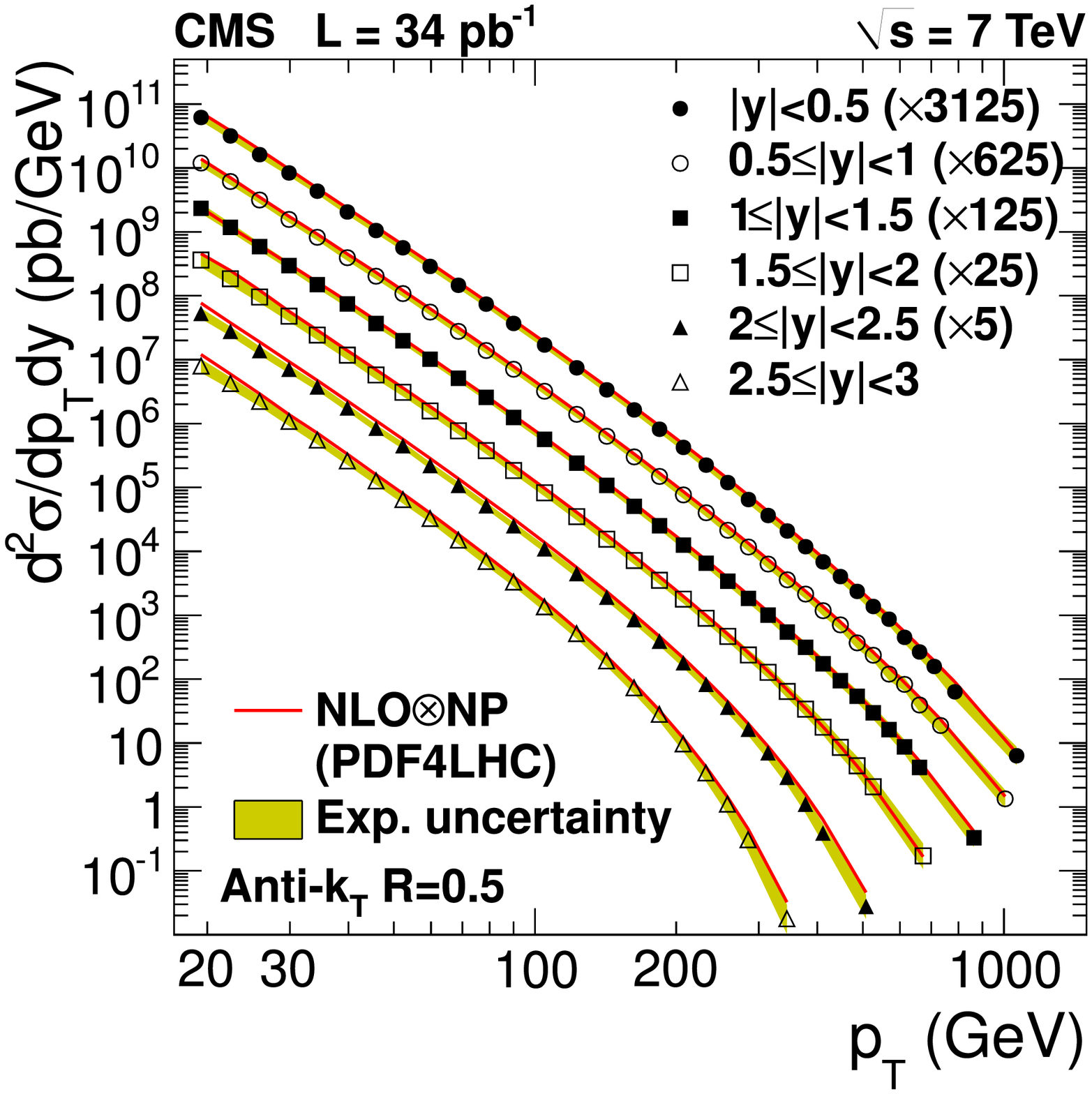} &
\includegraphics[height=17pc]{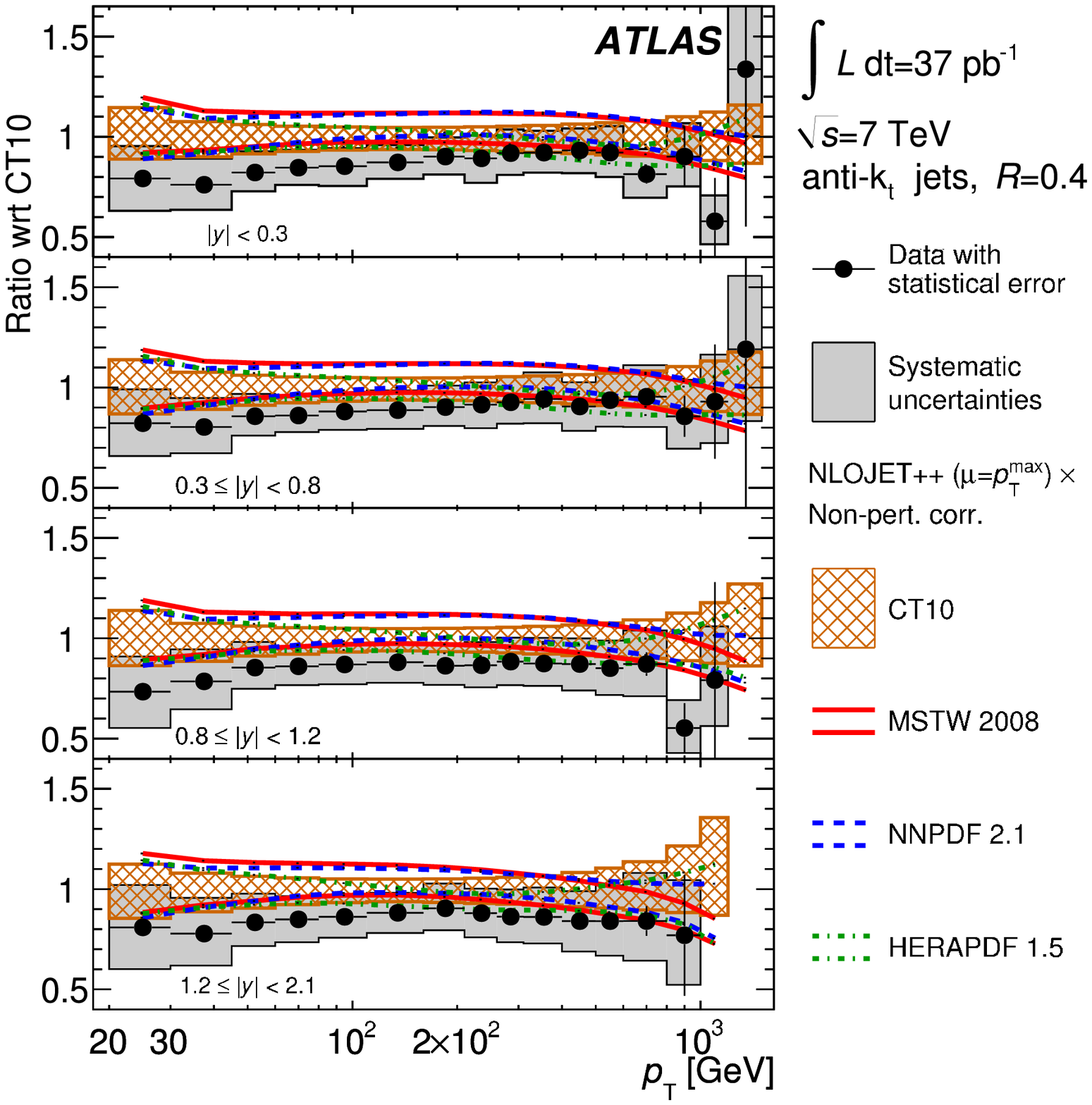}
\end{tabular}	
\end{center}
\caption{Measurements of the double-differential inclusive jet cross
  section, as a function of jet $p_T$ and rapidity.  The left plot
  shows the spectra as obtained by CMS \cite{:2011mea}, the right plot
  displays the ratio of the ATLAS measurements \cite{:2011fc} to the
  NLO prediction for different pdf sets.}
\label{fig:incljets}
\end{figure}

The simplest cross section, and the first to be
measured~\cite{:2010wv}, is the inclusive jet cross section.
``Inclusive'' implies that all jets passing the relevant kinematic
cuts are counted, regardless of other activity in a collision
event. Even with the very small data set available from the
summer of 2010, the measurements extended to 500~GeV, and subsequent
measurements using the full 2010 dataset~\cite{:2011fc,:2011mea} cover
the region from 20~GeV up to 1.5~TeV and rapidities in the range
$|y|<4.4$, thus probing a considerably larger phase space than
previously possible at the Tevatron and spanning approximately
$7\times 10^{-5} < x < 0.9$ in Bjorken $x$.  Over the full range,
NLO QCD calculations are in good agreement with the
data (Fig.\ \ref{fig:incljets}), and there is sensitivity to the value
of $\alpha_s$ and to the parton distributions.

The above measurements make use of information from both the
charged-particle tracker and the calorimeters of the experiments, and
are thus sensitive to charged and (most) neutral energy. Jets have
also been measured using only charged
particles~\cite{Aad:2011gn,Chatrchyan:2011id}.  While this gives an
incomplete picture of the jet, the generally better resolution of
track measurements at low momentum does allow the jet momenta to be
measured to lower values. This allows the transition from soft to hard
physics to be studied, as the jets emerge from the more common low
$p_T$ scatters. The data have been used to improve phenomenological
models of hadronisation and other non-perturbative features of hadron
physics.

As the LHC luminosity has grown, hard-scattering events have started
to be accompanied by increasing numbers of additional low-$p_T$
proton-proton interactions, a phenomenon known as pile-up.
Each jet's energy is then biased by depositions from the pile-up.
Information from tracking, which resolves separate pileup-vertices,
can help correct for this bias.
An alternative method examines all jets in the event, including
those at low $p_T$, and uses the median density of $p_T$ to determine
the amount of pile-up contamination~\cite{Cacciari:2007fd}. A given
hard jet's $p_T$ is then corrected by an amount equal to the product
of that density and the jet's area coverage in rapidity and
azimuth~\cite{Cacciari:2008gn}.

Measuring more than one jet in an event-by-event cross section brings
the ability to pick out specific kinematic configurations. Dijet cross
sections have been measured for dijet masses in a range between
$\sim70$ GeV and $5$ TeV~\cite{:2011fc,Chatrchyan:2011qta}.  
Angular distributions in dijet events, which are closely related to
the polar scattering angle in the parton-parton centre-of-mass frame,
are used to probe perturbative QCD predictions, but also to search for
deviations from the SM predictions such as contact interactions or
quark substructure~\cite{Khachatryan:2011as,Aad:2011aj}.
In some corners of phase space (such as large rapidity intervals
between low-$p_T$ dijets) standard perturbative techniques become
unreliable, and the data allow for powerful tests of new calculational
tools that are currently being developed.
In addition dijet configurations have been measured that are
sensitive to the presence of additional QCD radiation, such as the
azimuthal angle between
dijets~\cite{Khachatryan:2011zj,daCosta:2011ni} and the multiplicity
of additional jets in between a jet pair separated by large rapidity
\cite{Aad:2011jz}.

Hadronic event shapes represent a complementary approach for testing
jet production beyond the dijet regime.  These observables are
functions of the four momenta in the hadronic final state that
characterize the topology of an event's energy flow.  So far they have
been measured using jet momenta as inputs to the event-shape
calculation and compared to parton shower models with and without
matching to higher multiplicity matrix elements
\cite{Khachatryan:2011dx}. Using individual reconstructed particles,
such as charged tracks, as inputs to these observables should also
give sensitivity to non-perturbative effects, such as the underlying
event.

\subsection{Jets containing $b$-quarks}

A substantial fraction of jets at the LHC contain $b$-quarks, and such
jets are of particular interest since the $b$-quark mass of around 5~GeV
provides an additional hard scale, meaning some soft divergences are
naturally cut off in perturbative calculations, and also that
perturbatively summable logarithms in the ratio of the $b$ mass to other
scales in the event may occur.  In addition, $b$-jets are produced in
top decays and in several scenarios for physics beyond the SM, and QCD
contributions such as the
gluon to $b\bar{b}$ splitting vertex are a source of background for
processes such as $H\rightarrow b\bar{b}$.

The presence of $b$-quarks is tagged with various algorithms, based on
the reconstruction of secondary vertices from the decays of hadrons
containing $b$-quarks, the measurement of track impact parameters or the
identification of muons with a sizable transverse momentum with
respect to the jet axis. Depending on the algorithmic working point,
$b$-tagging efficiencies in the range between $30\%$ and $85\%$ are
typically obtained, with relative uncertainties of $5-10\%$.
The corresponding light-quark mistag rates range between a few per-mille and
$\sim15\%$, with relative uncertainties of $10-15\%$.

\begin{figure}[htbp]
\begin{center}
\begin{tabular}{lr}
\includegraphics[height=17pc]{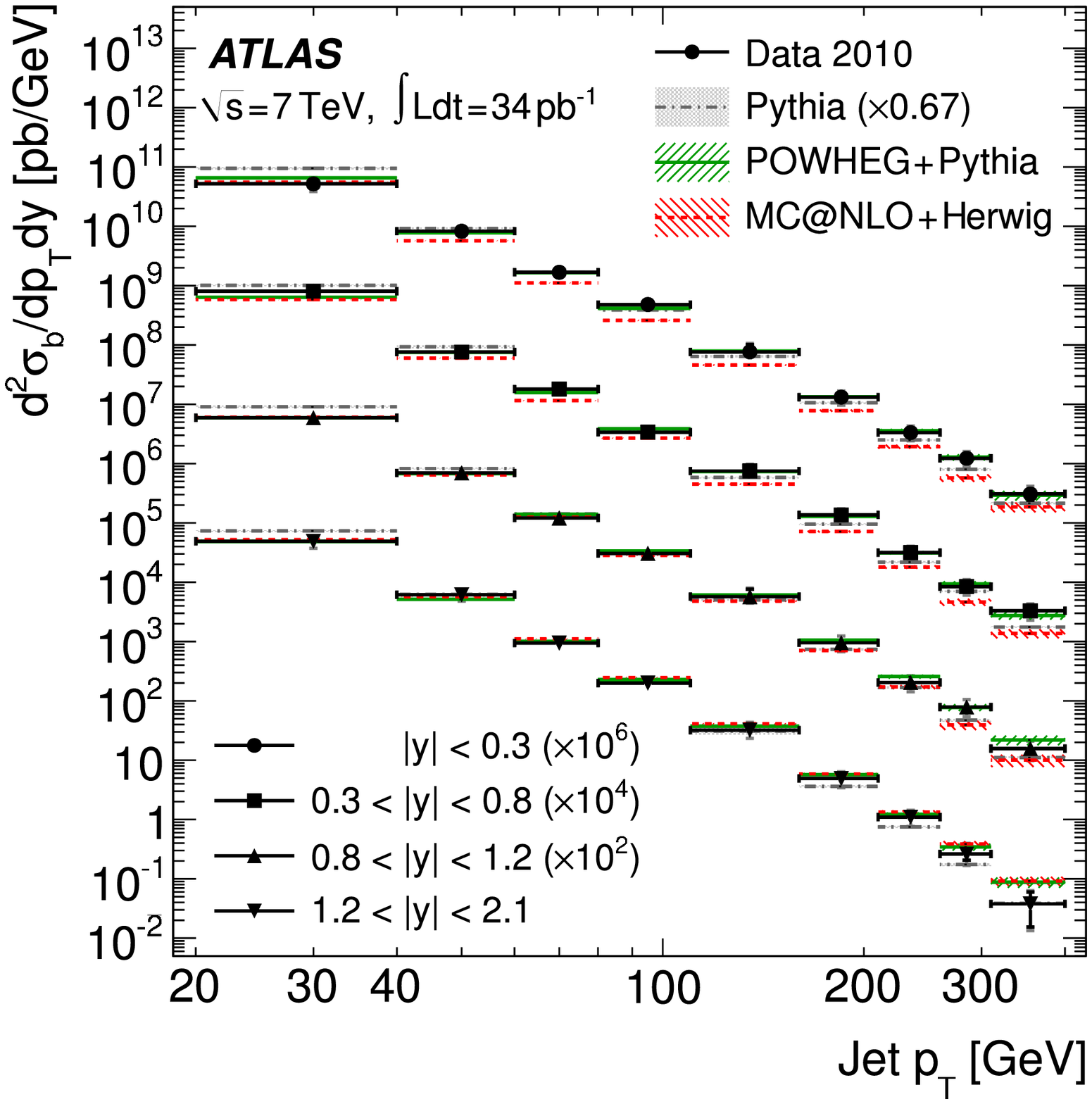} &
\includegraphics[height=17pc]{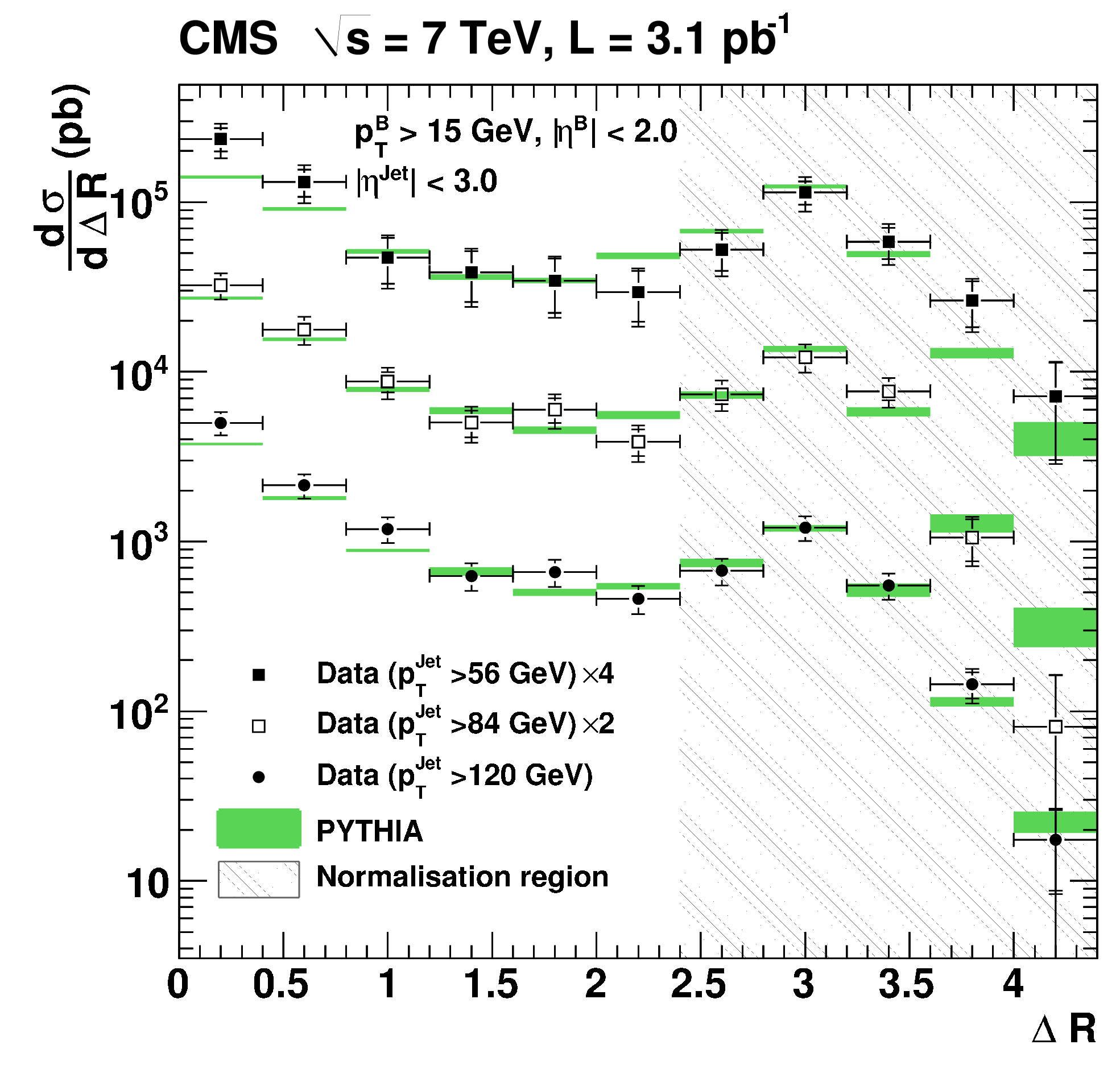}
\end{tabular}	
\end{center}
\caption{Left: Inclusive double-differential $b$-jet cross-section as a
  function of $p_T$ for different rapidity ranges, compared to the
  predictions of several Monte Carlo models \cite{ATLAS:2011ac};
  Right: Differential $B\bar{B}$ production cross section as a
  function of the angular separation 
  $\Delta R = [(y_B - y_{\bar B})^2 + (\phi_B - \phi_{\bar B})^2 ]^{\frac12}$ 
  (the Monte Carlo prediction is normalized to the region
  $\Delta R > 2.4$ (shaded)) \cite{Khachatryan:2011wq}.}
\label{fig:bjets}
\end{figure}

Measurements of pairs of such $b$-tagged jets have been made
\cite{ATLAS:2011ac} in the kinematic region $p_T > 40$ GeV and $|y| <
2.1$, with in addition the $b$-jets well separated in the azimuthal
plane (Fig.~\ref{fig:bjets} left). In this case the gluon splitting
$g\rightarrow b\bar{b}$ is expected to make a rather small
contribution. Note that these measurements have defined a $b$-jet as a
jet which is matched in angle to one or more $b$-hadrons.  NLO QCD
calculations describe the data well over the measured range.
Furthermore, measurements of the inclusive $b$-jet cross section have
been made over the range $20 < p_T < 400$ GeV and $|y| < 2.1$ and show
some discrepancies at higher rapidity $y$, as well as some divergence
between different NLO+PS calculations (POWHEG/MC@NLO), by up to 30\%.

An innovative study of the angular correlations between $b$-quarks
\cite{Khachatryan:2011wq} shows that a range of perturbative
calculations fail to describe the angular distribution of $b$-hadron
pairs in jet events (Fig.~\ref{fig:bjets} right). Specifically, when
normalized to the rate at wide angles, up to 50\% divergence between
data and theory, and between different approximations of QCD, is seen
at small angles.

The measurements to date suggest that a better understanding of the
$g\rightarrow b\bar{b}$ vertex may well be required in order to
accurately and correctly describe $b$-jet production over the kinematic
range accessible at the LHC.

Measurements of jets containing charm have also been
made~\cite{Aad:2011td}, using $D^*$-meson decays as a tag, but these
are much more sensitive to soft physics due to the lower $c$
mass. Significant discrepancies are seen between data and MC
simulations at low $z$, where $z$ is the fraction of the jet momentum
carried by the $D^*$.

\subsection{Jets produced in association with vector bosons}

The production of jets in association with a $W$ or $Z$ boson is
interesting for several reasons. The presence of a vector boson
introduces a hard scale, necessary for obtaining reliable fixed-order
perturbative calculations. Also, such processes are important
backgrounds for many other hard processes, such as top production, or
searches for supersymmetry or Higgs boson(s).  The vector bosons are
reconstructed in their leptonic ($e,\mu$) decay channels. Contrary to
inclusive jet studies, where backgrounds are not an issue, here some
backgrounds, such as from top quarks, have to be considered, in
particular in the $W$ case and at large jet multiplicities.  The
measurements \cite{Aad:2011qv,Collaboration:2012en,Chatrchyan:2011ne}
cover a phase space defined by the lepton acceptances as described
above and by jet momenta of $p_T > 30$ GeV and jet rapidities up to
$|y| < 4.4$.

\begin{figure}[htbp]
\begin{center}
\begin{tabular}{lr}
\includegraphics[height=17pc]{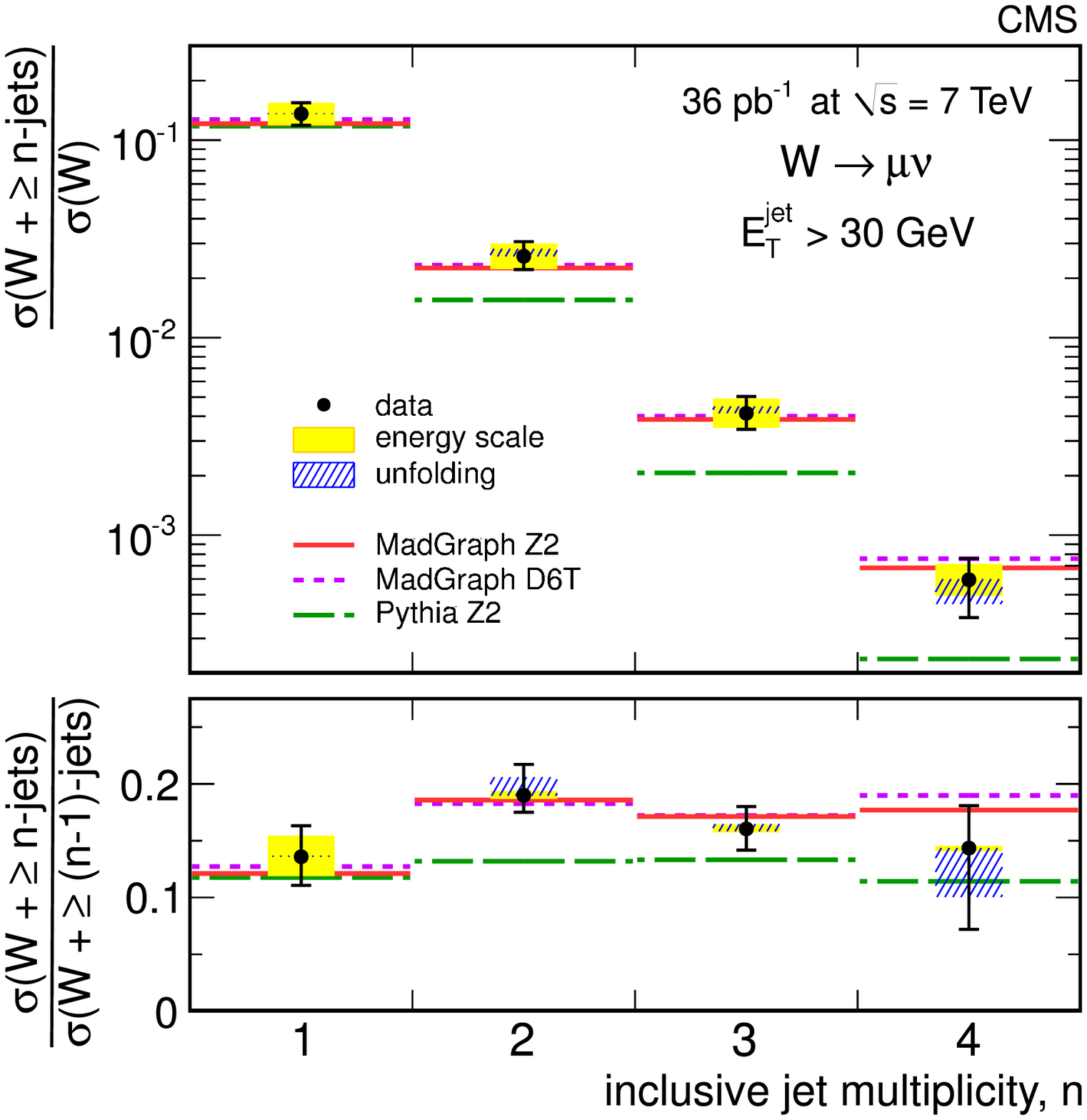} &
\includegraphics[height=17pc]{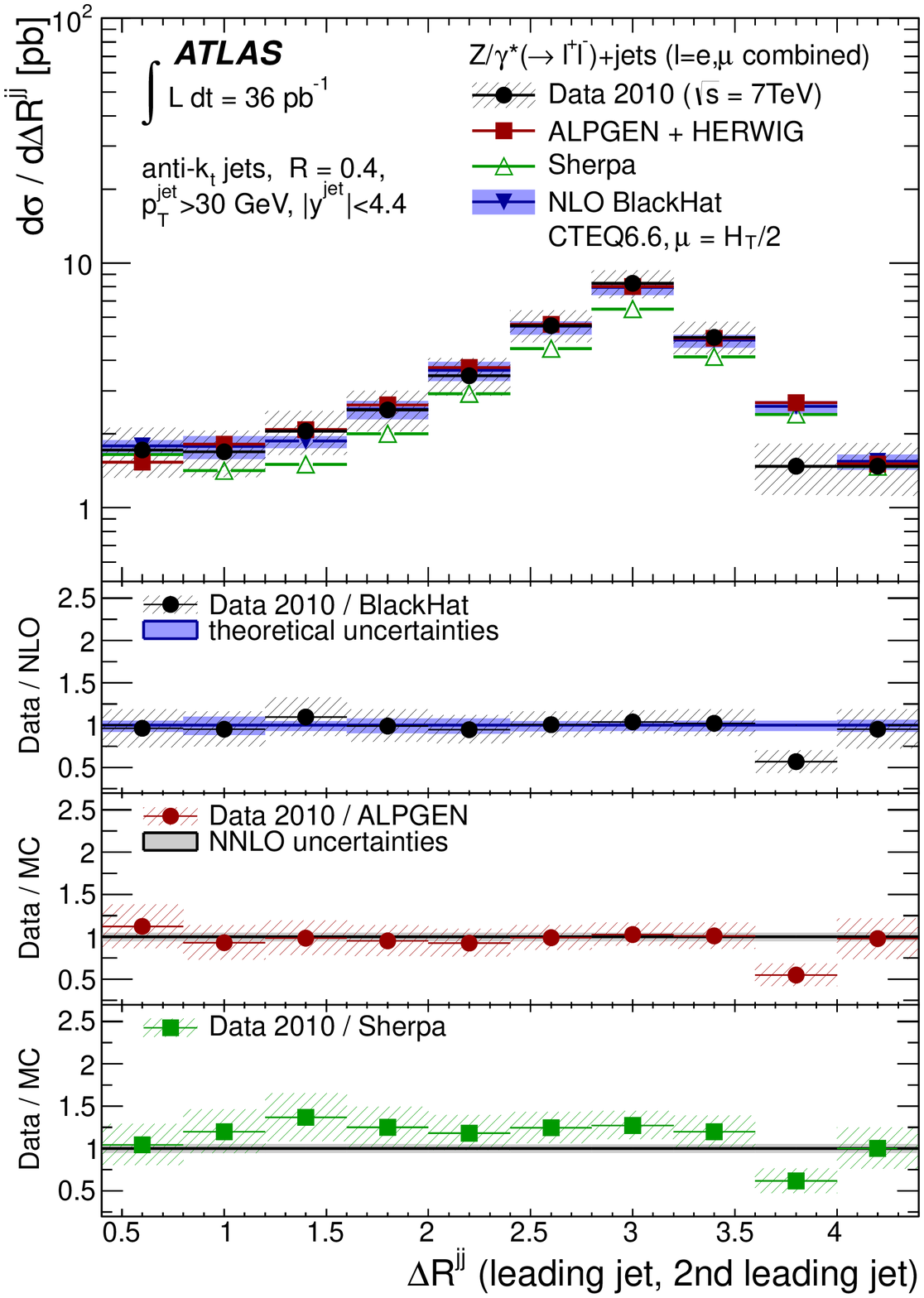}
\end{tabular}	
\end{center}
\caption{Left: Cross section for $W$ plus jets production, as a function
  of jet multiplicity and normalized to the inclusive $W$ cross section
  \cite{Chatrchyan:2011ne}; Right: Measured dijet cross section in
  $Z/\gamma^* (\rightarrow \ell\ell)+$jets production as a function of
  the angular separation, $\Delta R$, of the two leading jets
  \cite{Aad:2011qv}.}
\label{fig:V+jets}
\end{figure}

One important feature to emerge from the early LHC data is that to
correctly describe processes with a high multiplicity of high-$p_T$
jets, it is important to include higher multiplicity matrix elements
in the calculation.  This statement is on the face of it not
surprising, but in fact parton shower models matched simply to
two-to-two matrix elements had been remarkably successful in
describing a very wide variety of data at previous colliders.
At the LHC, there is clear evidence (such as the $H_T$ distribution in
multi-jet \cite{Aad:2011tq} and $W$ or $Z$ plus jets production) that the
parton shower description fails, generally providing a third jet which
is too hard, and/or too few hard jets overall.
One dramatic example is illustrated in Fig.~\ref{fig:V+jets} (left),
showing a factor of 2 discrepancy at high jet multiplicities with
Pythia (which showers $2\to1$ and $2\to 2$ configurations), and good
agreement to a prediction that includes matching of Pythia to multijet
tree-level matrix elements (here, MadGraph).

Similar observations are made for topological properties of such
events, such as angular distributions between the jets
(fig.~\ref{fig:V+jets} right). Also NLO calculations for up to 3 jets
in addition to the vector boson~\cite{Ellis:2009zw,Berger:2009ep} are
in good agreement with the data, an impressive and unprecedented
vindication of NLO QCD at high multiplicities.

The measurement of $b$-jets produced in association with $W$ or $Z$ bosons
provides another test of QCD in a multi-scale environment.
Measurements from the first year of LHC data
\cite{Aad:2011kp,Aad:2011jn,CMS-PAS-EWK-10-015} are consistent with
the expectations of NLO QCD, albeit with rather large uncertainties in
both theory and data. The cross section for $W$ plus $b$-jets lies above
the theory, as was also observed at the Tevatron, but more accuracy is
required before drawing any conclusion.  Finally, first measurements
of $W$ plus charm production are available \cite{CMS-PAS-EWK-11-013},
showing agreement with NLO QCD within the present accuracy.

\subsection{Jet Substructure and Boosted Heavy Particles}

The development of quarks and gluons into jets terminates in a
non-perturbative hadronisation process. However, before this low scale
is reached, large amounts of QCD radiation can take place at hard
momentum scales. This determines much of the jet's internal structure,
including the jet mass.  Measurement of such properties thus provides
another challenging point of comparison for perturbative QCD and an additional probe of hard
physics. Such measurements have in the past been used for example to
measure the strong coupling.

At the LHC, there is an additional reason for interest in jet
substructure.  Since by design the LHC opens up phase space well above
the electroweak symmetry-breaking scale, particles with masses about
this scale (e.g.\ $W$, $Z$, Higgs, top) can be produced well above
threshold, and therefore highly boosted. When such particles decay
to quarks, the quarks will be close in angle in the detector rest
frame, and thus the jets they give rise to may overlap or merge into a
single jet.

The momentum flow around the jet centre has been
measured~\cite{CMS-PAS-QCD-10-014,Aad:2011kq}
and shown to be reasonably well-modeled; charged particle
distributions within jets are likewise reasonably well
described~\cite{Aad:2011sc}, at least at high jet transverse momentum, by
calculations matching perturbative matrix elements, parton showers,
and non-perturbative models of hadronisation and underlying
event. Preliminary studies of new variables developed for searches at
the LHC~\cite{filtering,pruning} have been shown by both ATLAS and
CMS~\cite{Altheimer:2012mn}, and indicate that the promise of these variables,
including a reduced dependency on soft physics and pile-up, is borne
out. This area is still in its early stages and rapid progress is to
be expected over the next months.

\section{Photons}

Prompt, isolated photons provide a further look into the short
distance physics of high energy hadron collisions. Photons do not
undergo hadronisation and so, unlike quarks and gluons, can be
directly observed. However, they are also copiously produced from
secondary hadron decays (especially $\pi^0$ decays), and there are
subtleties associated with applying isolation criteria and suppressing
backgrounds, which require careful treatment. Measurement of photons
was a priority in the design of both experiments since the $H
\rightarrow \gamma\gamma$ channel is the most sensitive at low Higgs
masses.

\begin{figure}[htbp]
\begin{center}
\begin{tabular}{lr}
\includegraphics[height=17pc]{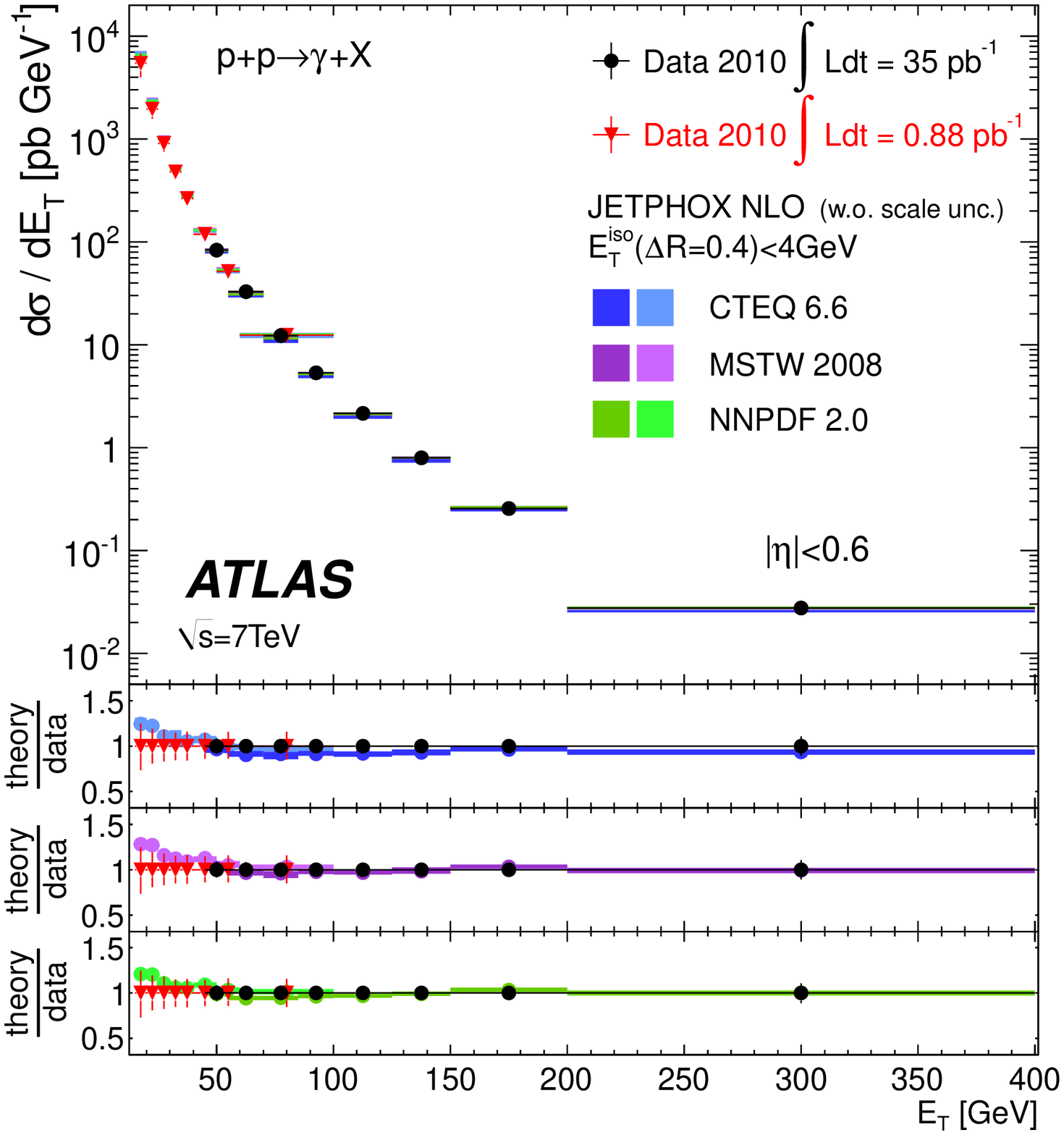} &
\includegraphics[height=17pc]{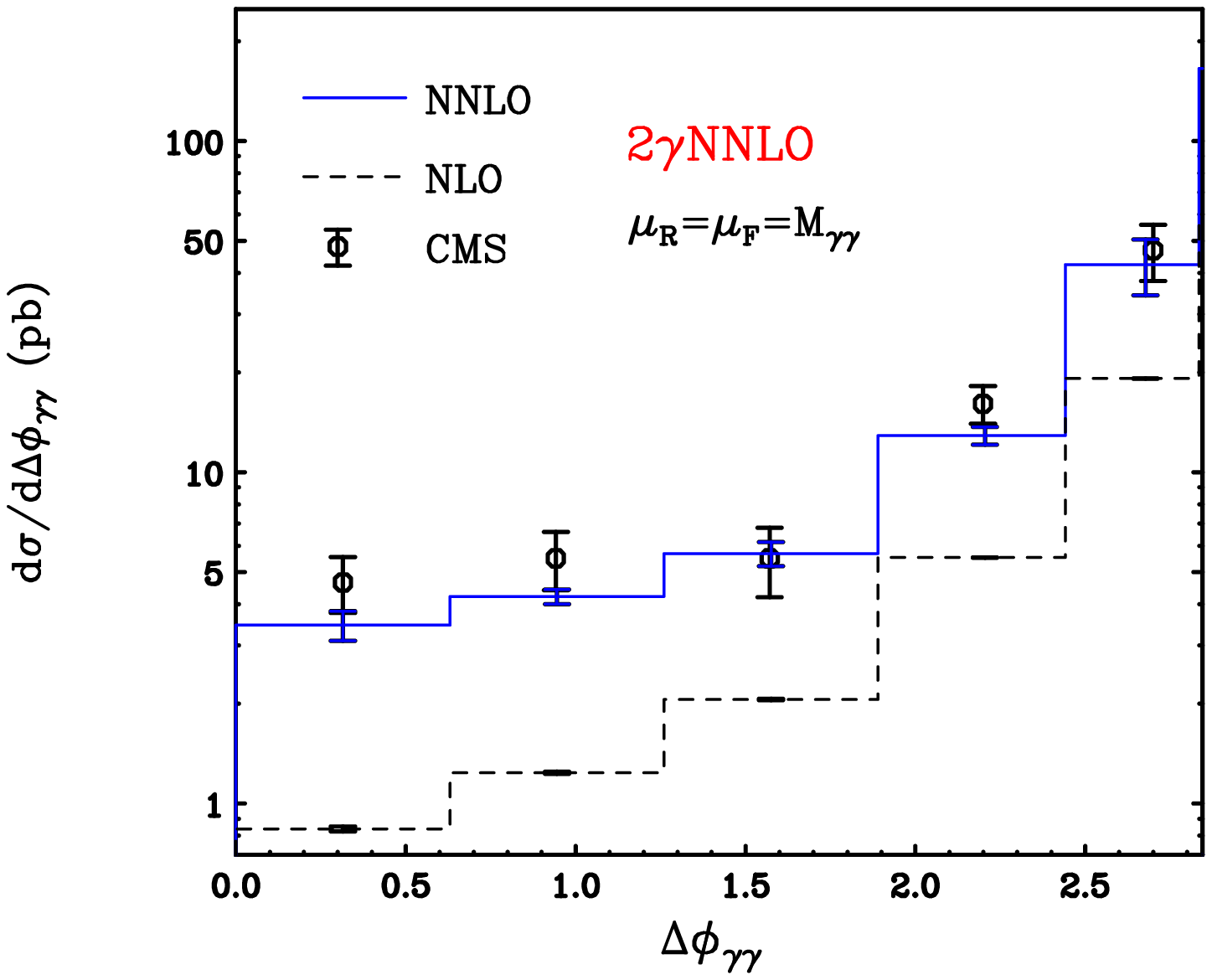}
\end{tabular}	
\end{center}
\caption{Left: Inclusive photon production cross section, as a
  function of photon $E_T$ \cite{ATL-PHYS-PUB-2011-013}; Right:
  Diphoton cross section as a function of the azimuthal separation of
  the two photons, with data from CMS~\cite{Chatrchyan:2011qt}
  compared to a recent NNLO calculation~\cite{Catani:2011qz} (plot
  from \cite{diphoton-NNLO-dphi}).}
\label{fig:photons}
\end{figure}

Fig.~\ref{fig:photons} (left) illustrates the precise measurement of
the inclusive photon cross section from 20~GeV to
400~GeV~\cite{ATL-PHYS-PUB-2011-013,Aad:2010sp,Aad:2011tw,Khachatryan:2010fm}.
To dynamically and reproducibly suppress the effects of pile-up and
underlying event on the photon isolation, the median/area techniques
described earlier for jets were for the first time applied also to
photons and shown to be successful.
The resulting measurements are in good agreement with the predictions
of NLO QCD.

Fig.~\ref{fig:photons} (right) shows a measurement of the diphoton
cross section as a function of the azimuthal angle between the
photons, $\Delta \phi_{\gamma\gamma}$~\cite{Chatrchyan:2011qt,Aad:2011mh},
compared to the recent calculation of~\cite{Catani:2011qz}.
It helps illustrate that there are sometimes regions of phase space
where even NLO calculations may fail dramatically. 
In this particular instance, the problem arises because in the region
of $\Delta \phi_{\gamma\gamma} < \pi$ the cross section is zero in
leading-order diphoton production, a simple consequence of momentum
conservation.
Thus NLO is actually the lowest non-zero order. NNLO additionally
introduces new topologies (e.g. $qq \to qq\gamma \gamma$), resulting
in large corrections and much better agreement with the data.

Continuing such measurements to higher precision is an essential
component of the Higgs search programme as well as the general
exploration of physics at the LHC.

\section{Weak Vector Bosons}

Measurements of jets produced in association with weak vector bosons
have already been discussed, above. More inclusive measurements,
focusing solely on the properties and kinematics of the bosons, also
provide important information.

\begin{figure}[htbp]
\begin{center}
\begin{tabular}{lr}
\includegraphics[height=12pc]{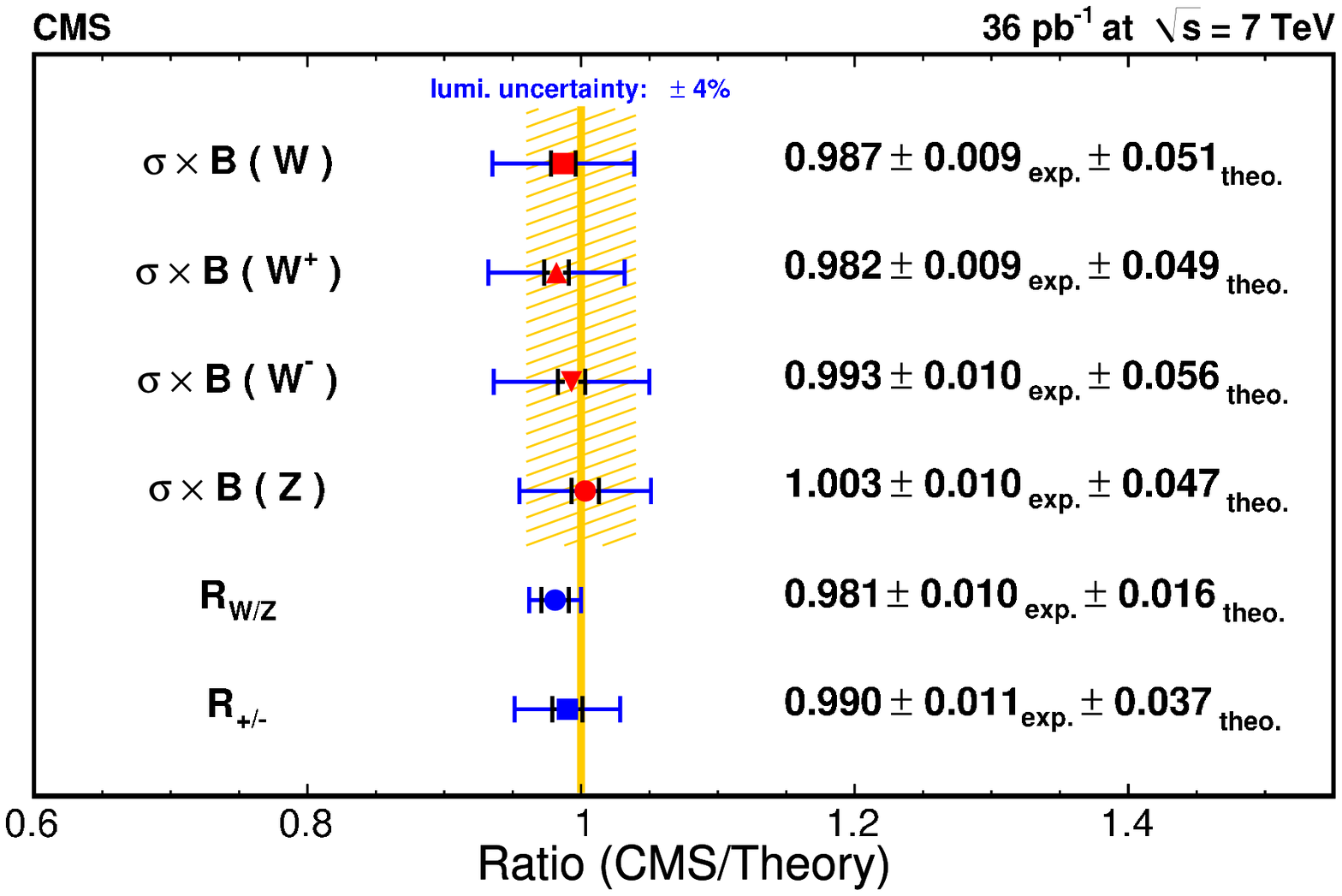} &
\includegraphics[height=17pc]{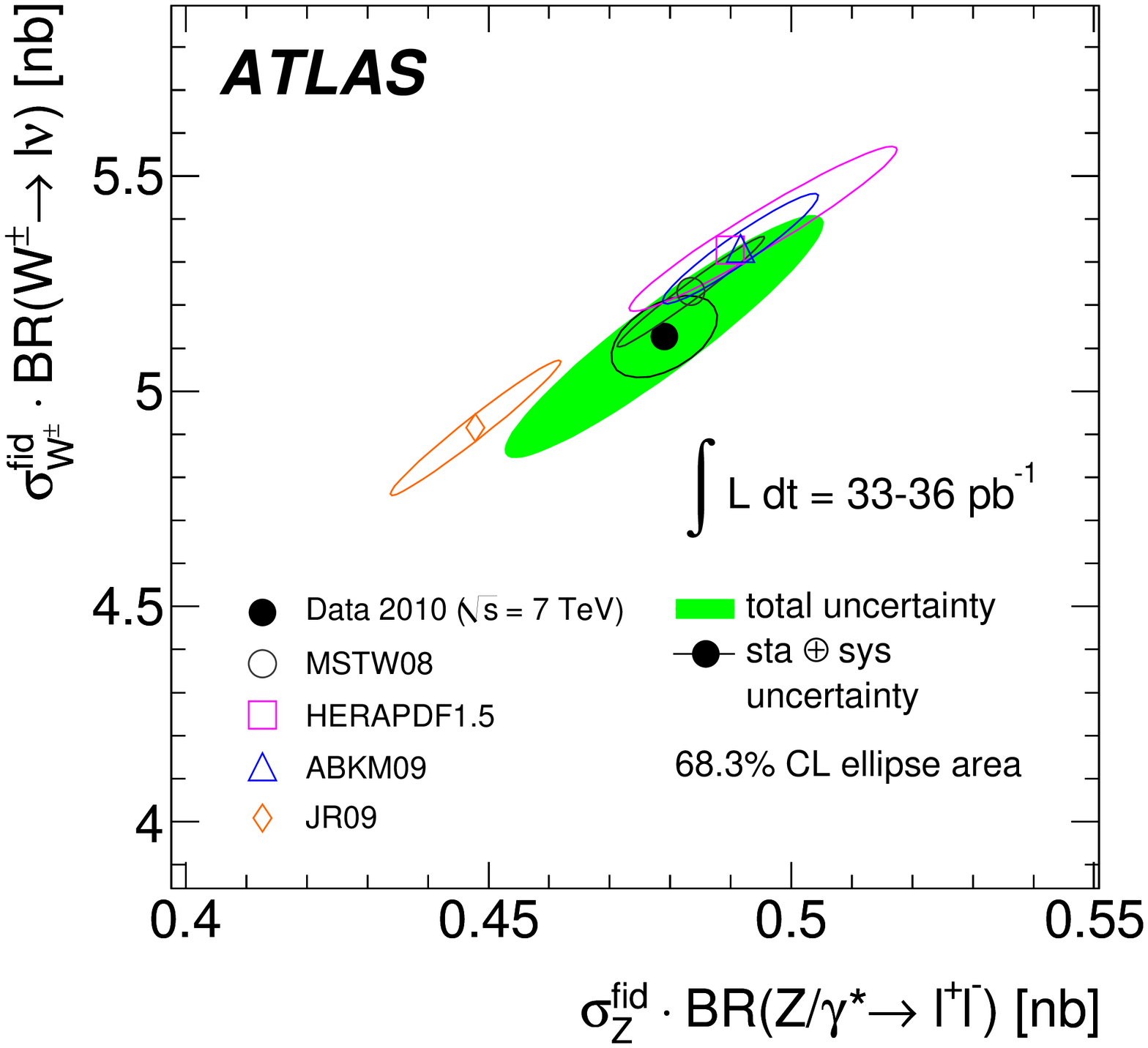}
\end{tabular}	
\end{center}
\caption{Inclusive $W$ and $Z$ boson production: Cross sections
  extrapolated to full phase space and compared to NNLO theory
  predictions \cite{CMS-Twiki-EWK} (left); Measured and predicted
  fiducial cross sections times leptonic branching ratios
  \cite{Aad:2011dm}, compared to predictions obtained using a variety
  of pdf sets (right).}
\label{fig:inclVbosons}
\end{figure}

In these cases, as also in the boson-plus-jet measurements, the
measurements are in fact of leptons and missing transverse momentum in
carefully defined regions reflecting the acceptance of the
detectors. This maximises the experimental accuracy and minimises the
model dependence of the results. The definition of a lepton is also
not entirely trivial, with the effects and treatment of QED radiation,
especially in electron channels, being as high as a few per cent in
some regions. While such effects are very precisely calculated for the
dominant soft and collinear contributions, wider-angle photon
production contains significant interference effects between initial
and final state radiation, and in principle triple gauge couplings, 
which are not so
well constrained. One solution~\cite{Butterworth:2010ym} adopted in
several measurements is to sum photons close to the leptons into the
lepton momentum, and apply no further correction for wider angle
photons. This provides an unambiguous definition of the measurement in
terms of final-state particles, reduces sensitivity to arbitrarily
soft and collinear photons (exploiting the good theoretical
understanding of these contributions) and avoids assumptions about the
less well known contributions. 
New theoretical
calculations~\cite{fewz,dynnlo} provide the ability to compare to NNLO
QCD in the actual phase-space region visible to the experiments,
avoiding the uncertainties associated with extrapolation to total
cross sections (Fig.~\ref{fig:inclVbosons}). 

\begin{figure}
  \centering
  \includegraphics[height=17pc]{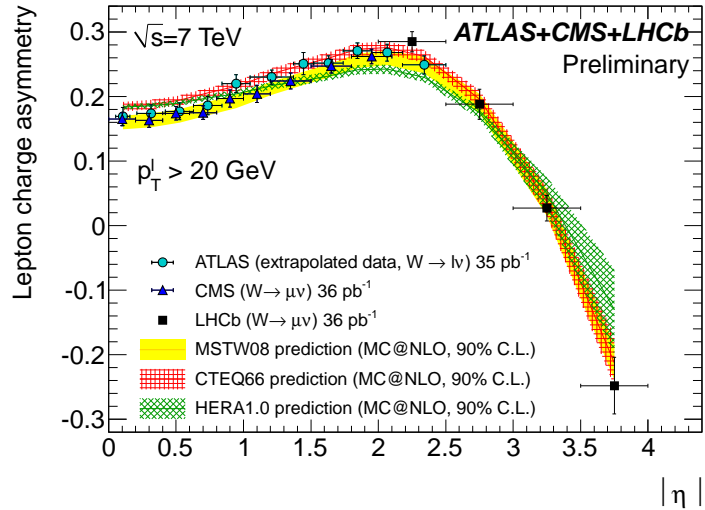}
  \caption{A comparison of ATLAS, CMS and LHCb results on the lepton
    charge asymmetry with theoretical predictions based on several
    different pdf sets~\cite{ATLAS-CONF-2011-129}. 
}
  \label{fig:lepton-charge-asymmetry}
\end{figure}

\begin{figure}[htbp]
\begin{center}
\begin{tabular}{lr}
\includegraphics[height=17pc]{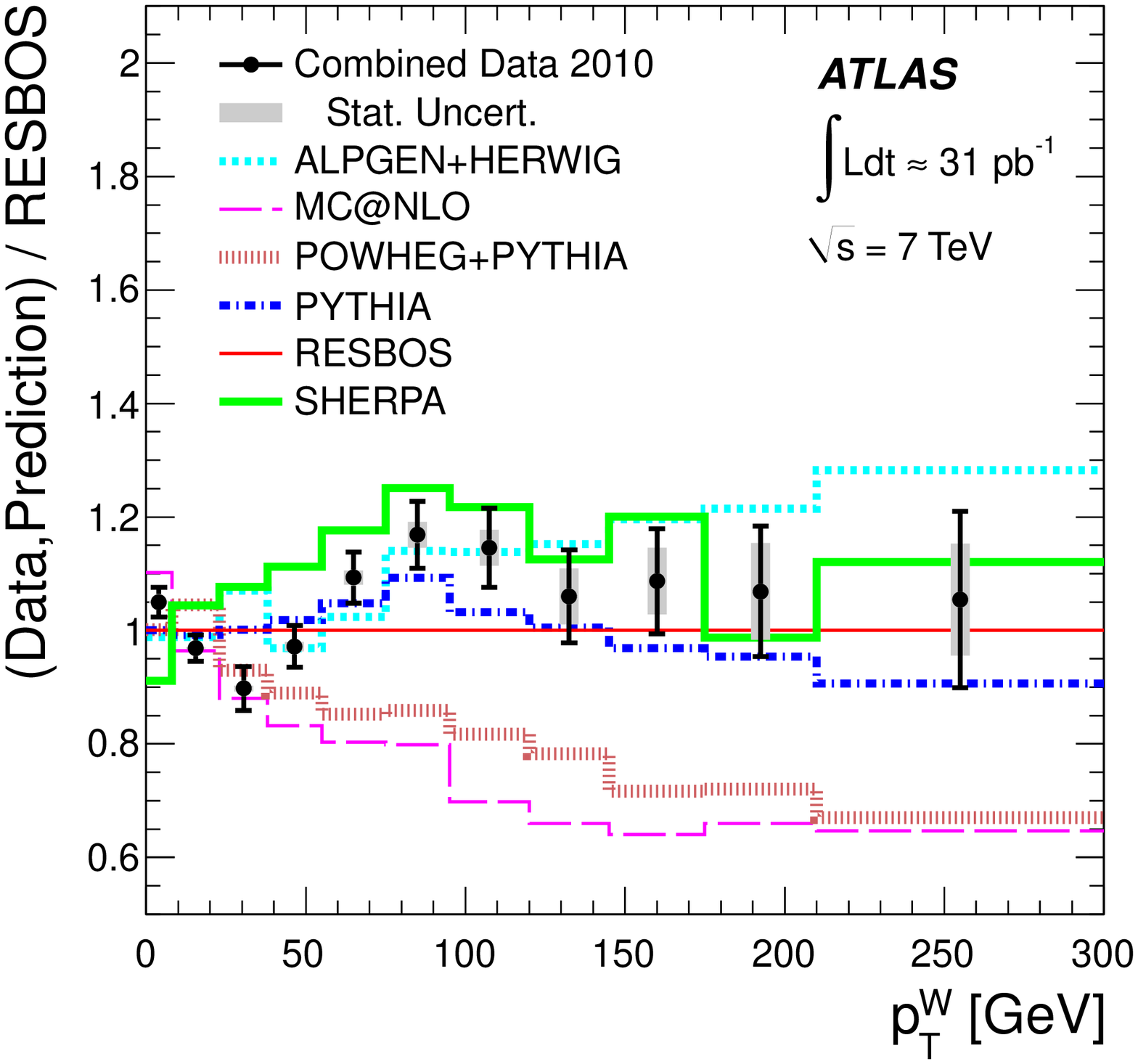}&%% $W$ pt v. MCs
\includegraphics[height=17pc]{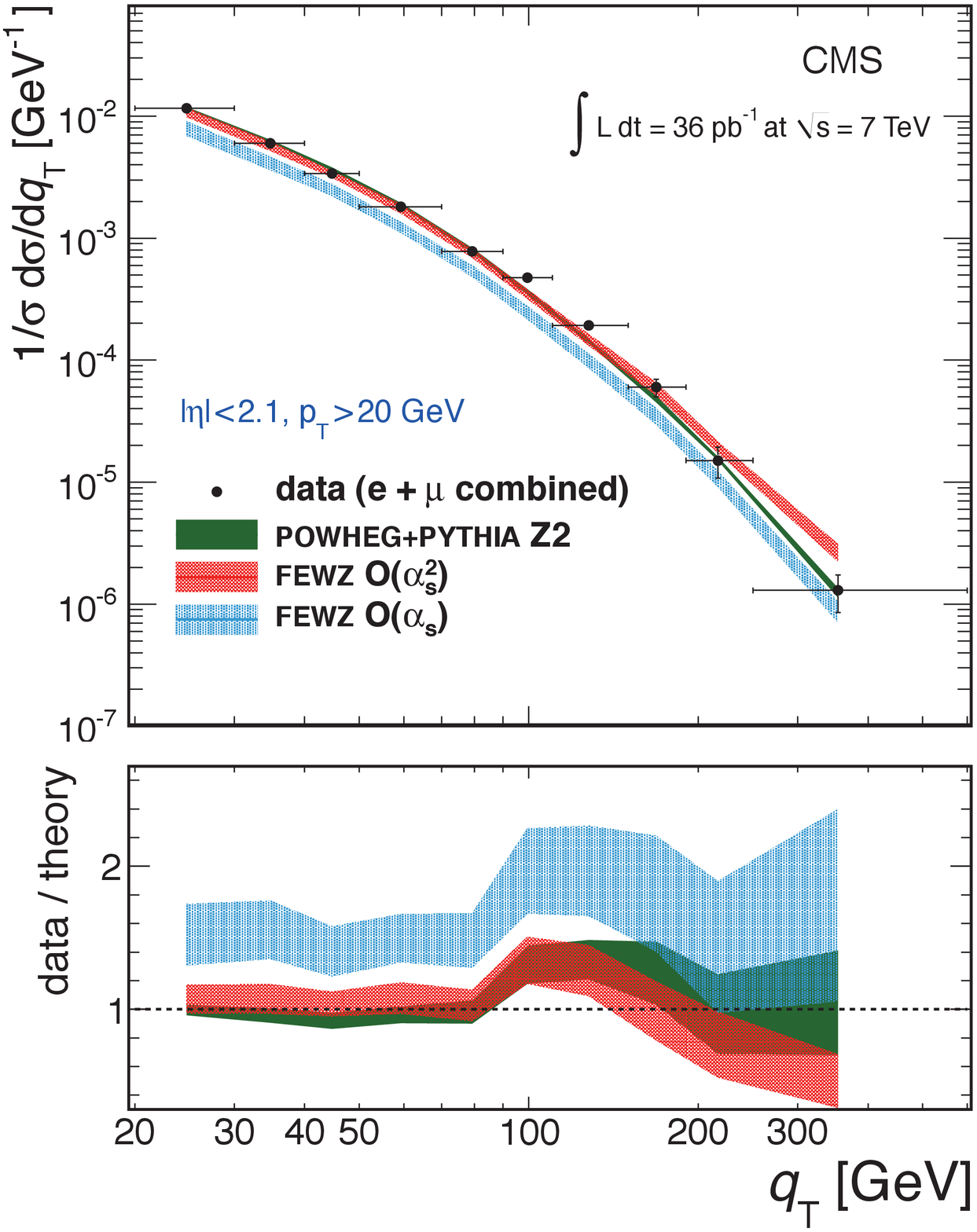}     %% $Z$ pt v. FO
\end{tabular}	
\end{center}
\caption{ATLAS \cite{Aad:2011fp} and CMS \cite{Chatrchyan:2011wt}
  results respectively for the $W$ and $Z$ $p_T$ distributions,
  compared to predictions from a range of theoretical
  tools. 
    }
\label{fig:WandZpt}
\end{figure}

The rapidity distributions of vector bosons in particular, and their
ratios, provide powerful constraints on the parton densities.  In
Fig.~\ref{fig:lepton-charge-asymmetry}, the ratio of positive to
negative lepton rapidity over the 
range accessible by ATLAS, CMS and LHCb is shown as an example.  
The $p_T$ distributions of the $W$~\cite{Aad:2011fp} and in particular
the $Z$ boson~\cite{Aad:2011gj,Chatrchyan:2011wt} constrain
initial-state QCD radiation at low $p_T$ and test
% multi-leg  -- not really as such
matrix element calculations at high $p_T$ (Fig.~\ref{fig:WandZpt}). They are
also an important precursor measurement to eventual $W$ mass and width
measurements, which remain the most important input to indirect
constraints on the Higgs mass and vital to checking the consistency of
the electroweak sector of the Standard Model. The $W$ to $e$ and $\mu$ channels provide 
competitive constraints on lepton universality. 

While the most precise and differential measurements come from the $e$
and $\mu$ channels, cross sections in $\tau$ decay channels have also
been measured~\cite{Aad:2011kt,Aad:2011fu,Chatrchyan:2011nv}, showing 
excellent consistency, and demonstrating the capability of the detectors to access
$\tau$ final states, which are important for many searches.

Finally, the
polarisation of $W$ bosons has been successfully
measured~\cite{Chatrchyan:2011ig}. Especially in vector-boson fusion,
this is a critical capability, since the Higgs mechanism provides the
longitudinal component of the $W$ and $Z$: measurement of final-state
polarisations will be fundamentally sensitive to production mechanisms
and to electroweak symmetry-breaking itself. Current measurements are
in good agreement with the Standard Model.

\section{Top Quarks}
\label{sec:top}
  
The measurement of top quark production and the determination of top
quark properties is interesting for several reasons. First of
all, the top is the heaviest quark, with a mass close to the
electroweak scale. This suggests that the top quark might play a
special role in electroweak symmetry breaking. Furthermore, in many
models for physics beyond the SM, the new states couple preferentially
to top quarks and might manifest themselves, eg., via decays into
$t\bar{t}$ pairs. Because of its very short lifetime
($\sim5\times10^{-25}$ sec), the top quark decays before it has a
chance to undergo hadronisation. This gives a rather unique access to
its properties (mass, charge, spin, couplings), in contrast to lighter
quarks which appear inside bound states. In particular, the top mass
plays an important role in many precision measurements, such as in
electroweak fits which constrain the allowed Higgs mass
range. Finally, besides representing a fundamental test of QCD
predictions, top quark production is an important background for new
physics searches, such as supersymmetry.
 
At the LHC, the dominant production mechanism is gluon fusion into a
$t\bar{t}$ pair ($\sim85\%$), while quark-antiquark annihilation
contributes the remaining $15\%$. This scenario, basically inverted
compared to the Tevatron, can be understood from the fact that at
$\sqrt{s}=7$ TeV a Bjorken $x$ scale of few $10^{-2}$ (where the gluon
density is dominant) is probed, while at the Tevatron the relevant $x$
range is $\sim2\times 10^{-1}$ (where quark densities are relatively more
important; furthermore the antiprotons provide an abundant supply of
antiquarks).
The top quark decays almost exclusively to a $W$ boson and a bottom
quark, which gives rises to a set of final states that are classified
according to the $W$ decays.  In about 5\% of the cases both $W$ bosons
decay to an electron or muon. This dilepton signature, accompanied by
$b$-quark jets and large missing energy due to the two neutrinos from
the $W$ decays, is interesting because it benefits from low background levels
(mostly $Z$+jets production). If one of the $W$s decays hadronically, we
have the lepton+jets signature, with a larger branching ratio
($\sim30\%$ for electron/muon decays), but also a somewhat larger
background, dominated by $W$ plus jets production. Finally, the all-hadronic
signature is the channel with the largest branching ratio
($\sim45\%$), but it suffers from a huge QCD multi-jet background. It
is clear that precision top studies rely on a complete understanding
of all detector components, since they involve precise jet, lepton and
MET reconstruction, as well as $b$-jet tagging.
 
\begin{figure}[htbp]
\begin{center}
\includegraphics[height=15pc]{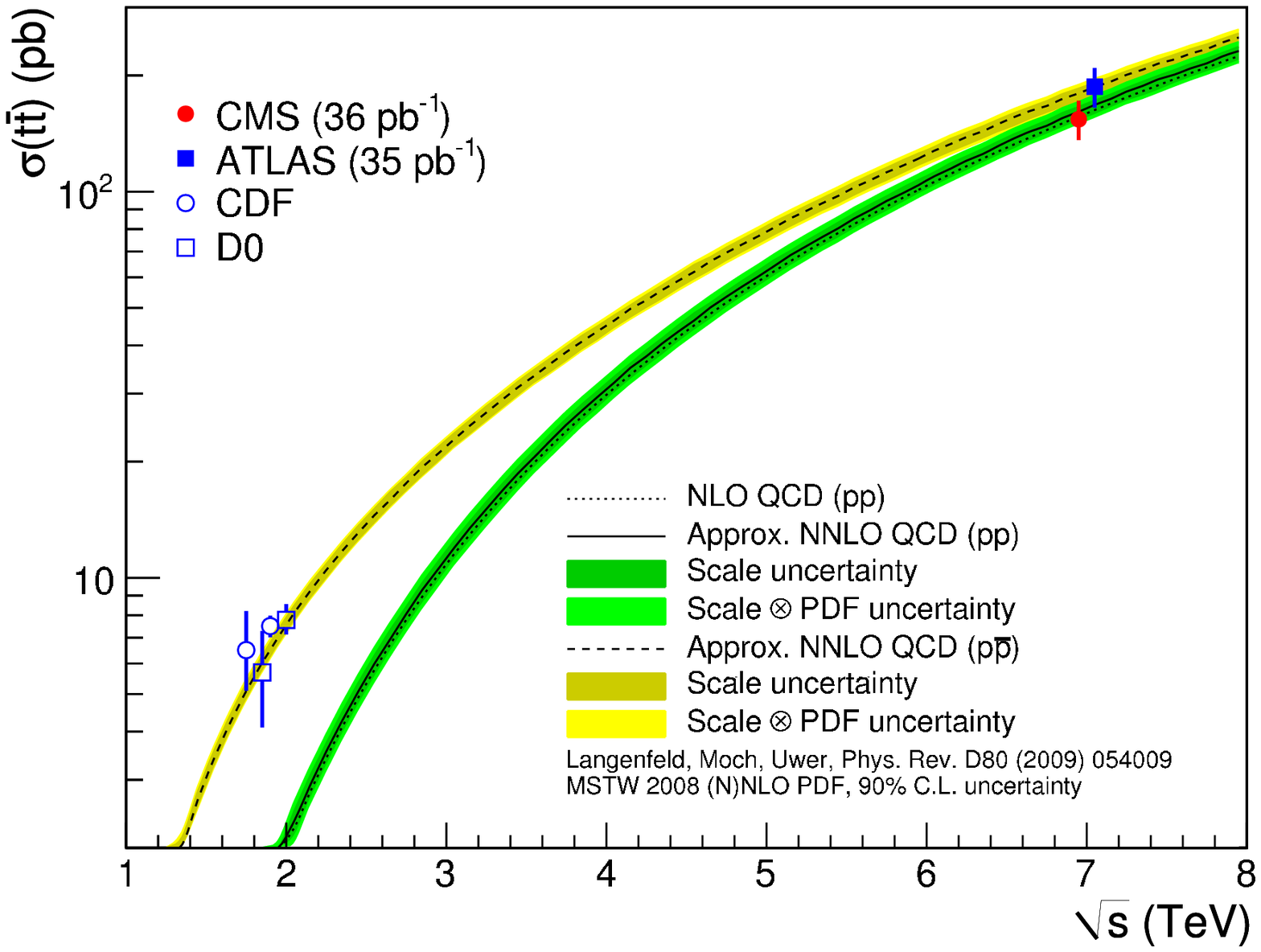}
\end{center}
\caption{Comparison of top-pair production cross sections as
predicted by higher-order QCD calculations and measured at the
Tevatron and the LHC \cite{Aad:2012qf,Chatrchyan:2011yy}; Plot adapted 
from \cite{CMS-TOP-Twiki}.}
\label{fig:top-results}
\end{figure}

By now, the LHC experiments have measured the top-pair production
cross section in almost all channels, even involving $\tau$ decays of the 
$W$. The most precise determinations are achieved in the lepton+jets
channel, where template distributions are fitted to the secondary
vertex mass as a function of the overall ($b$-tagged) jet
multiplicity. The vertex mass is an excellent discriminator for
tagging $b$-decays and for large jet multiplicities the top purity is
very high. From the simultaneous fit of the vertex mass in the various
jet/tag categories, not only the cross section is extracted, but also
the most relevant systematic uncertainties are constrained, such as
the jet energy scale, the $b$-tagging efficiency and the QCD scales
involved in the $W$+jets background simulation.  Based on the
2010 data, cross sections have been published
with a precision of about $12\%$ \cite{Aad:2012qf,Chatrchyan:2011yy},
and in preliminary results including 2011 data this has already been 
reduced to about $8\%$ \cite{CMS-PAS-TOP-11-024,ATLAS-CONF-2011-108}, 
which is approaching the
accuracy of approximate NNLO QCD predictions ($\sim7\%)$. The
central values  are in good agreement with the
predictions, cf.\ Fig.\ \ref{fig:top-results}, left. The convergence
of the proton-proton and antiproton-proton predictions at large
centre-of-mass energy, seen in Fig.\ \ref{fig:top-results} left, can
again be understood from pdf considerations, as above. At this
precision, the measurements start to become sensitive to differences
among the predictions based on various pdf sets. However, the
experimental results are already systematics limited, thus can only be
improved by a better control of the relevant parameters as mentioned
before. An interesting step towards cancellation of some of the
systematics, such as the luminosity uncertainty, consists in measuring
the ratio of the inclusive top-pair and $Z$ cross sections, in
particular in the dilepton channel. The pdf uncertainties in the
prediction for this ratio are anti-correlated, since there is a quark
pdf dominance for the inclusive $Z$ production, which should provide
further constraints on the available pdfs.
 
The next steps beyond the determination of the inclusive cross section
consist of differential measurements, such as top production in the
lepton+jets channel as a function of the (additional) jet multiplicity
\cite{ATLAS-CONF-2011-142}, or as a function of the $t\bar{t}$
invariant mass. The latter distribution is also a very sensitive tool
to search for new resonances decaying into $t\bar{t}$ pairs, for
example new heavy vector bosons, resonances in technicolour models,
heavy supersymmetric Higgs bosons or Kaluza-Klein graviton
excitations. At very high masses, with strongly boosted top quarks,
their decay products (jets and leptons) tend to be close by and merged
into fat jets. Here the recently developed ``top-tagging'' tools (see,
eg., \cite{Plehn:2011tg} for a review) start to be deployed by the LHC
experiments. First promising results \cite{CMS-PAS-EXO-11-006,Altheimer:2012mn} 
show that
it is indeed possible to resolve the internal structure of such fat
jets.  Finally, with larger statistics at the horizon, both in 2012
and then after the LHC shutdown in 2013-2014, a whole new group of
measurements will become feasible, by testing cross sections and
couplings in $t\bar{t}+X$ final states, where $X$ can be jets,
photons, vector bosons ($W$ and $Z$), missing transverse energy or
ultimately a Higgs boson.  First attempts in this direction exist
already, for example in the case of photons
\cite{ATLAS-CONF-2011-153}.
 
As with the cross section, the lepton+jets channel is also the
best signature for the top mass measurement, since it provides more
kinematic constraints than the dilepton channel with two
neutrinos. The top mass is reconstructed by combining the jets from
the hadronic $W$ decay with a further $b$-jet from the top decay. There
are several methods in order to obtain an optimal combination, reduced
combinatorics and best statistical sensitivity, such as the Ideogram
approach \cite{CMS-PAS-TOP-10-009} or a template fit
\cite{ATLAS-CONF-2011-120}.
%, as shown in Fig.\ \ref{fig:top-results}, right. 
The latter approach, giving a preliminary result of $m_\mathrm{top} = (175.9\pm
0.9\,\mathrm{(stat)} \pm 2.7\,\mathrm{(syst)})$ GeV/$c^2$ from a data
sample of $0.7$ fb$^{-1}$, already has a statistical uncertainty
comparable to be world's best result from the Tevatron ($173.2\pm 0.6
\pm 0.8$ GeV/$c^2$). The current systematic limitation is given by the
control of the $b$-jet energy scale. A rather complementary approach,
with very different systematic uncertainties and thus an important
cross check, consists in extracting the top quark mass from a
measurement of the cross section, utilizing the top mass dependence of
the perturbative QCD predictions. In this case the extracted mass is a
well defined parameter, eg., the running top quark mass in the
$\overline{\mathrm{MS}}$ scheme, which can then be converted into an
equivalent pole mass, in order to be compared to the direct
measurements mentioned before. In terms of precision, the currently
available results \cite{CMS-PAS-TOP-11-008, ATLAS-CONF-2011-054} are
not yet competitive with the direct reconstruction. The uncertainties
of $\sim7$ GeV/$c^2$ contain important contributions from systematics in the theory
predictions, related to the choice of renormalization/factorization
scales, of pdf sets and the strong coupling constant. 

First steps towards a detailed mapping of top quark properties have
been undertaken, such as a measurement of the top-antitop mass
difference as a direct test of CPT invariance
\cite{CMS-PAS-TOP-11-019}. Here the charge sign is determined from the
charge of the lepton in the reconstruction of the semi-leptonic top
decay.  Within the uncertainties of $1.3$ GeV$/c^2$ the result is
consistent with equal mass for the top and its antiparticle. The top
charge asymmetry is a further measurement relying on the separation of
top and antitop quarks, since it is obtained from the distribution of
the (pseudo-)rapidity difference of the two charge states. This study
has attracted particular attention because of deviations at the
$>3\,\sigma$ level found at the Tevatron in the top forward-backward
asymmetry, which could be explained by the possible exchange of new
particles in the production diagrams. While the same asymmetry is not
directly accessible at the LHC, with its symmetric proton-proton initial state,
the rapidity-dependence of the charge asymmetry also carries (reduced) sensitivity to such new
physics.  So far no deviations from the SM expectations have been
found at the LHC. Finally, first determinations of the top electric
charge, the $W$ polarization in top decays and of $t\bar{t}$ spin
correlations have been possible with the 2011 data, and will be
further pursued with more data coming in.

Contrary to the top pair production in strong interactions, single top
quarks are produced through electroweak interactions. Thanks to the
much larger cross sections ($\mathcal{O}(70\,\mathrm{pb}^{-1})$) at
the LHC compared to the Tevatron, this process has already been probed
with the 2010 and 2011 data at the $30\%$ precision level. These
measurements, so far consistent with SM predictions, are interesting
in their own right as tests of the SM, but also as tools for searching
for new physics, such as flavour changing neutral currents, heavy $W$
partners or charged Higgs bosons. In addition, the final states to be
reconstructed, and the large backgrounds to be controlled, are of very
similar nature to those in low-mass Higgs searches, thus several
relevant tools have been sharpened via such studies.

\section{Dibosons}

The lowest cross section SM processes so far probed at the
LHC are those in which two electroweak vector bosons are
produced. These processes receive contributions from triple, and in
principle quartic, gauge boson couplings, and so are directly
sensitive to the gauge symmetry structure sector of the SM, as well as producing topologies characteristic of many
extensions of the SM. Diphotons (already discussed),
$W^+W^-$ and $ZZ$ production are also key Higgs boson search channels.

The cross sections for $W\gamma, Z\gamma$~\cite{Aad:2011tc,Chatrchyan:2011rr}, 
$WW$~\cite{Aad:2011kk,Chatrchyan:2011tz}, $ZZ$~\cite{Aad:2011xj} and 
$WZ$~\cite{Aad:2011cx} production
have all been measured. In most cases the visible cross section in a
well-defined phase space region has been measured,
as well as
an extrapolation to the total cross section (except for those
involving photons, where a $p_T$ cut on the photon, at around 20~GeV, is
still required in order to define a physical final state). For several
channels, limits have been derived on triple-gauge couplings in various
constrained approximations. 

Since the top quark decays to $W$ bosons, a particular challenge for measurements of 
$WW$ production is the background from top production, either in pairs or singly 
in association with a $W$. 
To suppress this, a veto is often applied on the presence of jets, especially $b$-jets. 
However, there is a strong interest in measuring diboson production in the presence of jets, 
especially since vector-boson fusion or scattering processes have in general two jets in 
the final state. These processes, which have an even lower cross section
than diboson production via parton fusion, are another key search channel and studying them 
will be essential to validate any picture of electroweak symmetry breaking or other new physics
which emerges from Higgs searches and other studies.

Studies of dibosons are thus only just beginning. 
More statistics are required to measure
differential cross sections and fully characterise these processes, and the couplings, in a 
less model-dependent fashion. 
Nevertheless, even with published datasets of 1~fb$^{-1}$ or less, the LHC results
are competitive with LEP and Tevatron data.

%======================================================================
\section{Conclusions}

\begin{figure}[htbp]
\centerline{
\includegraphics[height=24pc]{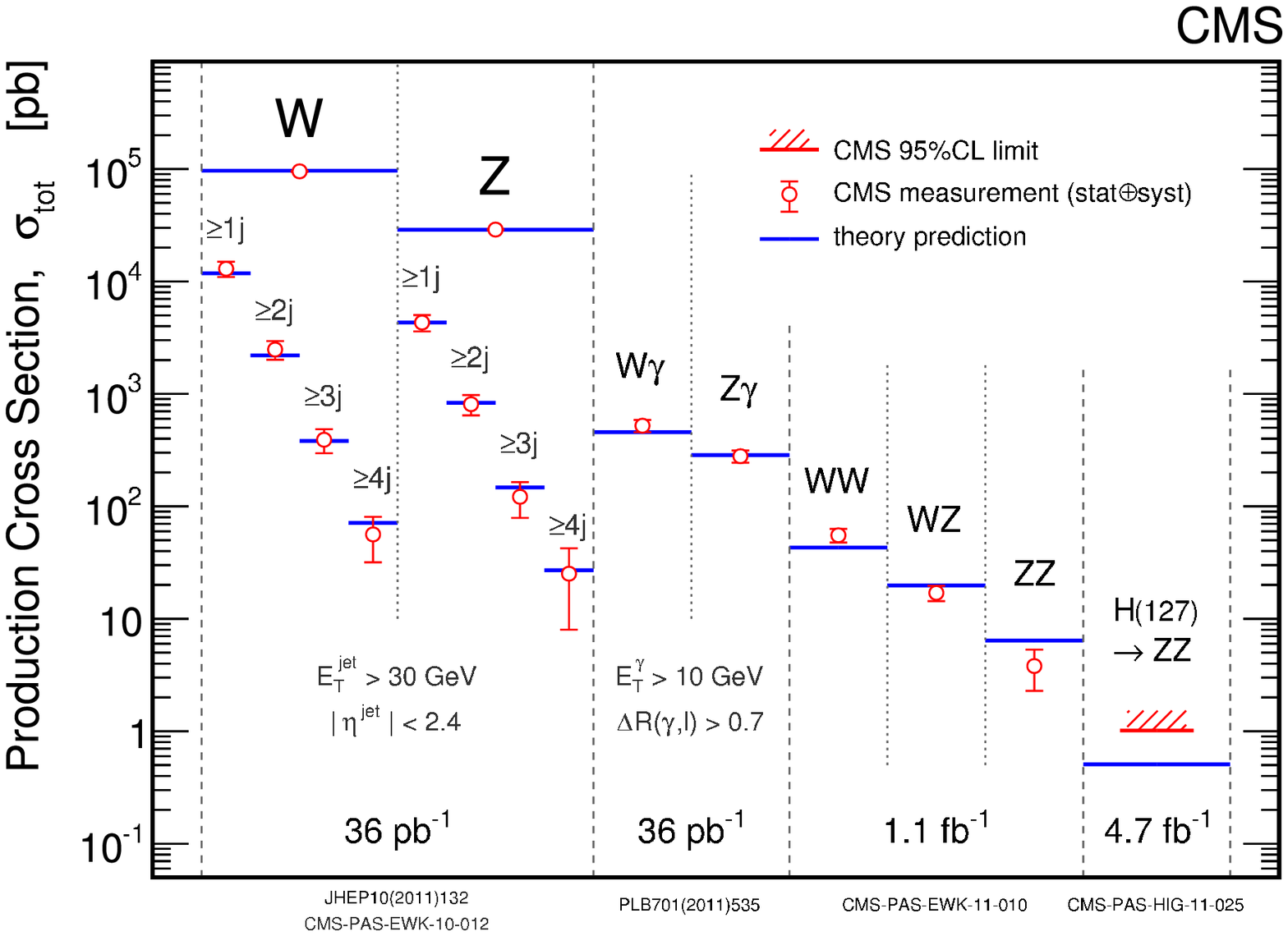} 
}
\caption{Summary of cross sections for hard processes measured at the
  LHC, where CMS results are used as an example \cite{CMS-Twiki-EWK}. 
Similar plots are available for ATLAS~\cite{ATLAS-summary}.}

\label{fig:summary-plot}
\end{figure}

Less than two years after the start of 7~TeV pp collisions, the LHC
experiments have delivered a broad array of analyses of hard
processes.
Experimental uncertainties in the measurements range from about 5\%,
for example for the $W$ and $Z$ boson cross sections, to about 20\%
for measurements of differential jet rates.
In nearly all cases there is excellent agreement with theoretical
predictions, especially those including higher-order corrections
and/or matching of fixed-order and parton-shower calculations, which
tend to have precisions comparable to experimental results.
A summary of data--theory comparisons is given in
Fig.~\ref{fig:summary-plot}.

This early success of the LHC program builds on many factors: the
impressive performance of the accelerator; the dedication of the large
numbers of experimenters within the collaborations; and also the
considerable preparatory work that took place in the years preceding
the LHC's startup. The latter involved understanding the detectors'
characteristics and calibration, as well as the development of a range
of tools for predicting the properties of collisions and for
performing sound comparisons between theory and data.

The study of hard processes at the LHC is not merely about comparing
data and theoretical predictions, but also about constraining our
knowledge of the standard model and its phenomenology.
This has already taken place for example in the context of the tuning
of Monte Carlo programs.
In the near future one can expect the LHC data to start providing
important constraints on pdfs and on fundamental electroweak
parameters such as the top-quark mass.
And the handful of observables that show discrepancies relative to
predictions, notably those sensitive to the presence of heavy flavor,
will hopefully spur the development of yet better predictive tools.

Finally, the overall good agreement between data and theoretical
predictions provides a solid foundation in the search for the Higgs
boson and physics beyond the standard model.
Indeed as analysis of the most recent data continues and as new data
are collected in the coming years, the LHC is becoming sensitive to
the full range of allowed Higgs-boson masses and it will start to
probe the region above the electroweak scale in a wide variety of
channels.
It is in part thanks to the studies of hard processes carried out so
far that this program of research can be pursued with confidence.

\section*{Acknowledgements}

The results discussed here were only possible because of the hard work 
and dedication of many thousands of people working on the LHC and its 
detectors. JMB and GD have particularly benefited from working closely 
with many colleagues
on ATLAS and CMS, and all three authors gratefully acknowledge this, as well
as discussions with the wider experimental and theoretical communities.
We additionally thank F.~P.~Schilling for assistance in replotting 
the top cross section. 
Despite all of this, responsibility for any errors or omissions is, 
of course, our own.
Finally, GPS wishes to acknowledge support from the French Agence Nationale de
la Recherche under grant ANR-09-BLAN-0060 and the European Commission
under grant PITN-GA-2010-264564
and JMB wishes to acknowledge the STFC, and the Royal Society for a 
Wolfson Research Merit award.

%%% Numbered Literature Cited

%% Caution: Not all Annual Reviews series use this format for
%% bibliography entries.Your Production Editor will advise you
%% on correct format for your particular series.


\begin{thebibliography}{99}
%\cite{Aad:2008zzm}
\bibitem{Aad:2008zzm}
  G.~Aad {\it et al.}  [ATLAS Collaboration],
  %``The ATLAS Experiment at the CERN Large Hadron Collider,''
  JINST {\bf 3} (2008) S08003.
  %%CITATION = JINST,3,S08003;%%

%\cite{:2008zzk}
\bibitem{:2008zzk}
  R.~Adolphi {\it et al.}  [CMS Collaboration],
  %``The CMS experiment at the CERN LHC,''
  JINST {\bf 3} (2008) S08004.
  %%CITATION = JINST,3,S08004;%%
  
 %\cite{Alves:2008zz}
\bibitem{Alves:2008zz}
  A.~A.~Alves, Jr. {\it et al.}  [LHCb Collaboration],
  %``The LHCb Detector at the LHC,''
  JINST {\bf 3} (2008) S08005.
  %%CITATION = JINST,3,S08005;%%
  
%\cite{Aamodt:2008zz}
\bibitem{Aamodt:2008zz}
  K.~Aamodt {\it et al.}  [ALICE Collaboration],
  %``The ALICE experiment at the CERN LHC,''
  JINST {\bf 3} (2008) S08002.
  %%CITATION = JINST,3,S08002;%%  
  
\bibitem{Alekhin:2011sk}
  S.~Alekhin, S.~Alioli, R.~D.~Ball, V.~Bertone, J.~Blumlein, M.~Botje, J.~Butterworth and F.~Cerutti {\it et al.},
  %``The PDF4LHC Working Group Interim Report,''
  arXiv:1101.0536 [hep-ph].

\bibitem{Lai:2010vv}
  H.~-L.~Lai, M.~Guzzi, J.~Huston, Z.~Li, P.~M.~Nadolsky, J.~Pumplin and C.~-P.~Yuan,
  %``New parton distributions for collider physics,''
  Phys.\ Rev.\ D {\bf 82} (2010) 074024
  [arXiv:1007.2241 [hep-ph]].
  %%CITATION = ARXIV:1007.2241;%%


\bibitem{Martin:2009iq}
  A.~D.~Martin, W.~J.~Stirling, R.~S.~Thorne and G.~Watt,
  %``Parton distributions for the LHC,''
  Eur.\ Phys.\ J.\ C {\bf 63} (2009) 189
  [arXiv:0901.0002 [hep-ph]].
  %%CITATION = ARXIV:0901.0002;%%

\bibitem{Ball:2011uy}
  R.~D.~Ball {\it et al.}  [The NNPDF Collaboration],
  %``Unbiased global determination of parton distributions and their uncertainties at NNLO and at LO,''
  Nucl.\ Phys.\ B {\bf 855} (2012) 153
  [arXiv:1107.2652 [hep-ph]].
  %%CITATION = ARXIV:1107.2652;%%

%\cite{Sjostrand:2006za}
\bibitem{Sjostrand:2006za}
  T.~Sjostrand, S.~Mrenna and P.~Z.~Skands,
  %``PYTHIA 6.4 Physics and Manual,''
  JHEP {\bf 0605} (2006) 026
  [hep-ph/0603175].
  %%CITATION = HEP-PH/0603175;%%

\bibitem{Bahr:2008pv}
  M.~Bahr, S.~Gieseke, M.~A.~Gigg, D.~Grellscheid, K.~Hamilton, O.~Latunde-Dada, S.~Platzer and P.~Richardson {\it et al.},
  %``Herwig++ Physics and Manual,''
  Eur.\ Phys.\ J.\ C {\bf 58} (2008) 639
  [arXiv:0803.0883 [hep-ph]].
  %%CITATION = ARXIV:0803.0883;%%

\bibitem{Gleisberg:2008ta}
  T.~Gleisberg, S.~.Hoeche, F.~Krauss, M.~Schonherr, S.~Schumann, F.~Siegert and J.~Winter,
  %``Event generation with SHERPA 1.1,''
  JHEP {\bf 0902} (2009) 007
  [arXiv:0811.4622 [hep-ph]].
  %%CITATION = ARXIV:0811.4622;%%

\bibitem{Alwall:2007fs}
  J.~Alwall, S.~Hoche, F.~Krauss, N.~Lavesson, L.~Lonnblad, F.~Maltoni, M.~L.~Mangano and M.~Moretti {\it et al.},
  %``Comparative study of various algorithms for the merging of parton showers and matrix elements in hadronic collisions,''
  Eur.\ Phys.\ J.\ C {\bf 53} (2008) 473
  [arXiv:0706.2569 [hep-ph]].
  %%CITATION = ARXIV:0706.2569;%%

\bibitem{Mangano:2002ea}
  M.~L.~Mangano, M.~Moretti, F.~Piccinini, R.~Pittau and A.~D.~Polosa,
  %``ALPGEN, a generator for hard multiparton processes in hadronic collisions,''
  JHEP {\bf 0307} (2003) 001
  [hep-ph/0206293].
  %%CITATION = HEP-PH/0206293;%%

\bibitem{Alwall:2011uj}
  J.~Alwall, M.~Herquet, F.~Maltoni, O.~Mattelaer and T.~Stelzer,
  %``MadGraph 5 : Going Beyond,''
  JHEP {\bf 1106} (2011) 128
  [arXiv:1106.0522 [hep-ph]].
  %%CITATION = ARXIV:1106.0522;%%

\bibitem{Frixione:2002ik}
  S.~Frixione and B.~R.~Webber,
  %``Matching NLO QCD computations and parton shower simulations,''
  JHEP {\bf 0206} (2002) 029
  [hep-ph/0204244].
  %%CITATION = HEP-PH/0204244;%%

%\cite{Nason:2004rx}
\bibitem{Nason:2004rx}
  P.~Nason,
  %``A New method for combining NLO QCD with shower Monte Carlo algorithms,''
  JHEP {\bf 0411} (2004) 040
  [hep-ph/0409146].
  %%CITATION = HEP-PH/0409146;%%

  %\cite{Dissertori:2003pj}
\bibitem{Dissertori:2003pj}
  G.~Dissertori, I.~G.~Knowles and M.~Schmelling,
  %``High energy experiments and theory,''
  (International series of monographs on physics. 115)
  
%\cite{Ellis:1991qj}
\bibitem{Ellis:1991qj}
  R.~K.~Ellis, W.~J.~Stirling and B.~R.~Webber,
  %``QCD and collider physics,''
  Camb.\ Monogr.\ Part.\ Phys.\ Nucl.\ Phys.\ Cosmol.\  {\bf 8} (1996) 1.
  %%CITATION = CMPCE,8,1;%%  
  
%\cite{Nakamura:2010zzi}
\bibitem{Nakamura:2010zzi}
  K.~Nakamura {\it et al.}  [Particle Data Group Collaboration],
  %``Review of particle physics,''
  J.\ Phys.\ G G {\bf 37} (2010) 075021.
  %%CITATION = JPHGB,G37,075021;%%  

\bibitem{Buckley:2011ms}
  A.~Buckley, J.~Butterworth, S.~Gieseke, D.~Grellscheid, S.~Hoche, H.~Hoeth, F.~Krauss and L.~Lonnblad {\it et al.},
  %``General-purpose event generators for LHC physics,''
  Phys.\ Rept.\  {\bf 504} (2011) 145
  [arXiv:1101.2599 [hep-ph]].
  %%CITATION = ARXIV:1101.2599;%%

%\cite{Cacciari:2008gp}
\bibitem{Cacciari:2008gp}
  M.~Cacciari, G.~P.~Salam and G.~Soyez,
  %``The anti-k_t jet clustering algorithm,''
  JHEP {\bf 0804} (2008) 063
  [arXiv:0802.1189 [hep-ph]].
  %%CITATION = JHEPA,0804,063;%%

%\cite{Chatrchyan:2011ds}
\bibitem{Chatrchyan:2011ds}
  S.~Chatrchyan {\it et al.}  [CMS Collaboration],
  %``Determination of Jet Energy Calibration and Transverse Momentum Resolution in CMS,''
  JINST {\bf 6} (2011) P11002
  [arXiv:1107.4277 [physics.ins-det]].
  %%CITATION = ARXIV:1107.4277;%%
  
%\cite{:2011he}
\bibitem{:2011he}
  G.~Aad {\it et al.} [ATLAS Collaboration],
  %``Jet energy measurement with the ATLAS detector in proton-proton collisions at sqrt(s) = 7 TeV,''
  arXiv:1112.6426 [hep-ex].
  %%CITATION = ARXIV:1112.6426;%%  

%\cite{:2010wv}
\bibitem{:2010wv}
  G.~Aad {\it et al.}  [ATLAS Collaboration],
  %``Measurement of inclusive jet and dijet cross sections in proton-proton collisions at 7 TeV centre-of-mass energy with the ATLAS detector,''
  Eur.\ Phys.\ J.\ C {\bf 71} (2011) 1512
  [arXiv:1009.5908 [hep-ex]].
  %%CITATION = ARXIV:1009.5908;%%

%\cite{:2011fc}
\bibitem{:2011fc}
  G.~Aad {\it et al.} [ATLAS Collaboration],
  %``Measurement of inclusive jet and dijet production in pp collisions at sqrt(s) = 7 TeV using the ATLAS detector,''
  arXiv:1112.6297 [hep-ex].
  %%CITATION = ARXIV:1112.6297;%%
  
  %\cite{:2011mea}
\bibitem{:2011mea}
  S.~Chatrchyan {\it et al.}  [CMS Collaboration],
  %``Measurement of the Inclusive Jet Cross Section in pp Collisions at sqrt(s) = 7 TeV,''
  Phys.\ Rev.\ Lett.\  {\bf 107} (2011) 132001
  [arXiv:1106.0208 [hep-ex]].
  %%CITATION = ARXIV:1106.0208;%%

%\cite{Aad:2011gn}
\bibitem{Aad:2011gn}
  G.~Aad {\it et al.}  [ATLAS Collaboration],
  %``Properties of jets measured from tracks in proton-proton collisions at center-of-mass energy sqrt(s) = 7 TeV with the ATLAS detector,''
  Phys.\ Rev.\ D {\bf 84} (2011) 054001
  [arXiv:1107.3311 [hep-ex]].
  %%CITATION = ARXIV:1107.3311;%%
  
%\cite{Chatrchyan:2011id}
\bibitem{Chatrchyan:2011id}
  S.~Chatrchyan {\it et al.}  [CMS Collaboration],
  %``Measurement of the Underlying Event Activity at the LHC with $\sqrt{s}= 7$ TeV and Comparison with $\sqrt{s} = 0.9$ TeV,''
  JHEP {\bf 1109} (2011) 109
  [arXiv:1107.0330 [hep-ex]].
  %%CITATION = ARXIV:1107.0330;%%

\bibitem{Cacciari:2007fd}
  M.~Cacciari and G.~P.~Salam,
  %``Pileup subtraction using jet areas,''
  Phys.\ Lett.\ B {\bf 659} (2008) 119
  [arXiv:0707.1378 [hep-ph]].
  %%CITATION = ARXIV:0707.1378;%%

\bibitem{Cacciari:2008gn}
  M.~Cacciari, G.~P.~Salam and G.~Soyez,
  %``The Catchment Area of Jets,''
  JHEP {\bf 0804} (2008) 005
  [arXiv:0802.1188 [hep-ph]].
  %%CITATION = ARXIV:0802.1188;%%
    

%\cite{Chatrchyan:2011qta}
\bibitem{Chatrchyan:2011qta}
  S.~Chatrchyan {\it et al.}  [CMS Collaboration],
  %``Measurement of the differential dijet production cross section in proton-proton collisions at sqrt(s)=7 TeV,''
  Phys.\ Lett.\ B {\bf 700} (2011) 187
  [arXiv:1104.1693 [hep-ex]].
  %%CITATION = ARXIV:1104.1693;%%
  
%\cite{Khachatryan:2011as}
\bibitem{Khachatryan:2011as}
  V.~Khachatryan {\it et al.}  [CMS Collaboration],
  %``Measurement of Dijet Angular Distributions and Search for Quark Compositeness in pp Collisions at $sqrt{s} = 7$ TeV,''
  Phys.\ Rev.\ Lett.\  {\bf 106} (2011) 201804
  [arXiv:1102.2020 [hep-ex]].
  %%CITATION = ARXIV:1102.2020;%%
  
  %\cite{Aad:2011aj}
\bibitem{Aad:2011aj}
  G.~Aad {\it et al.}  [ATLAS Collaboration],
  %``Search for New Physics in Dijet Mass and Angular Distributions in pp Collisions at $\sqrt{s} = 7$ TeV Measured with the ATLAS Detector,''
  New J.\ Phys.\  {\bf 13} (2011) 053044
  [arXiv:1103.3864 [hep-ex]].
  %%CITATION = ARXIV:1103.3864;%%  

%\cite{Khachatryan:2011zj}
\bibitem{Khachatryan:2011zj}
  V.~Khachatryan {\it et al.}  [CMS Collaboration],
  %``Dijet Azimuthal Decorrelations in $pp$ Collisions at $\sqrt{s} = 7$~TeV,''
  Phys.\ Rev.\ Lett.\  {\bf 106} (2011) 122003
  [arXiv:1101.5029 [hep-ex]].
  %%CITATION = ARXIV:1101.5029;%%
  
  %\cite{daCosta:2011ni}
\bibitem{daCosta:2011ni}
  G.~Aad {\it et al.}  [ATLAS Collaboration],
  %``Measurement of Dijet Azimuthal Decorrelations in pp Collisions at sqrt(s)=7 TeV,''
  Phys.\ Rev.\ Lett.\  {\bf 106} (2011) 172002
  [arXiv:1102.2696 [hep-ex]].
  %%CITATION = ARXIV:1102.2696;%%

%\cite{Aad:2011jz}
\bibitem{Aad:2011jz}
  G.~Aad {\it et al.}  [ATLAS Collaboration],
  %``Measurement of dijet production with a veto on additional central jet activity in pp collisions at sqrt(s)=7 TeV using the ATLAS detector,''
  JHEP {\bf 1109} (2011) 053
  [arXiv:1107.1641 [hep-ex]].
  %%CITATION = ARXIV:1107.1641;%%

%\cite{Khachatryan:2011dx}
\bibitem{Khachatryan:2011dx}
  V.~Khachatryan {\it et al.}  [CMS Collaboration],
  %``First Measurement of Hadronic Event Shapes in pp Collisions at sqrt(s)=7 TeV,''
  Phys.\ Lett.\ B {\bf 699} (2011) 48
  [arXiv:1102.0068 [hep-ex]].
  %%CITATION = ARXIV:1102.0068;%%

%\cite{ATLAS:2011ac}
\bibitem{ATLAS:2011ac}
  G.~Aad {\it et al.}  [ATLAS Collaboration],
  %``Measurement of the inclusive and dijet cross-sections of b-jets in pp collisions at sqrt(s) = 7 TeV with the ATLAS detector,''
  Eur.\  Phys.\  J.\  C {\bf 71} (2011) 1846
  [arXiv:1109.6833 [hep-ex]].
  %%CITATION = ARXIV:1109.6833;%%

%\cite{Khachatryan:2011wq}
\bibitem{Khachatryan:2011wq}
  V.~Khachatryan {\it et al.}  [CMS Collaboration],
  %``Measurement of B anti-B Angular Correlations based on Secondary Vertex Reconstruction at sqrt(s)=7 TeV,''
  JHEP {\bf 1103} (2011) 136
  [arXiv:1102.3194 [hep-ex]].
  %%CITATION = ARXIV:1102.3194;%%

\bibitem{Aad:2011td}
  G.~Aad {\it et al.}, [ATLAS Collaboration],
  %``Measurement of D*+/- meson production in jets from pp collisions at sqrt(s) = 7 TeV with the ATLAS detector,''
  arXiv:1112.4432 [hep-ex].
  %%CITATION = ARXIV:1112.4432;%%

%\cite{Aad:2011qv}
\bibitem{Aad:2011qv}
  G.~Aad {\it et al.}, [ATLAS Collaboration],
  %``Measurement of the production cross section for Z/gamma* in association with jets in pp collisions at sqrt(s) = 7 TeV with the ATLAS detector,''
  arXiv:1111.2690 [hep-ex].
  %%CITATION = ARXIV:1111.2690;%%
  
%\cite{Collaboration:2012en}
\bibitem{Collaboration:2012en}
  G.~Aad {\it et al.}  [ATLAS Collaboration],
  %``Study of jets produced in association with a W boson in pp collisions at sqrt(s) = 7 TeV with the ATLAS detector,''
  arXiv:1201.1276 [hep-ex].
  %%CITATION = ARXIV:1201.1276;%%

%\cite{Chatrchyan:2011ne}
\bibitem{Chatrchyan:2011ne}
  S.~Chatrchyan {\it et al.}  [CMS Collaboration],
  %``Jet Production Rates in Association with W and Z Bosons in pp Collisions at sqrt(s) = 7 TeV,''
  arXiv:1110.3226 [hep-ex].
  %%CITATION = ARXIV:1110.3226;%%

%\cite{Aad:2011tq}
\bibitem{Aad:2011tq}
  A.~Collaboration {\it et al.}  [ATLAS Collaboration],
  %``Measurement of multi-jet cross sections in proton-proton collisions at a 7 TeV center-of-mass energy,''
  Eur.\ Phys.\ J.\ C {\bf 71} (2011) 1763
  [arXiv:1107.2092 [hep-ex]].
  %%CITATION = ARXIV:1107.2092;%%

\bibitem{Ellis:2009zw}
  R.~K.~Ellis, K.~Melnikov and G.~Zanderighi,
  %``Generalized unitarity at work: first NLO QCD results for hadronic $W^+$ 3jet production,''
  JHEP {\bf 0904} (2009) 077
  [arXiv:0901.4101 [hep-ph]].
  %%CITATION = ARXIV:0901.4101;%%

\bibitem{Berger:2009ep}
  C.~F.~Berger, Z.~Bern, L.~J.~Dixon, F.~Febres Cordero, D.~Forde, T.~Gleisberg, H.~Ita and D.~A.~Kosower {\it et al.},
  %``Next-to-Leading Order QCD Predictions for W+3-Jet Distributions at Hadron Colliders,''
  Phys.\ Rev.\ D {\bf 80} (2009) 074036
  [arXiv:0907.1984 [hep-ph]].
  %%CITATION = ARXIV:0907.1984;%%

%\cite{Aad:2011kp}
\bibitem{Aad:2011kp}
  G.~Aad {\it et al.}  [ATLAS Collaboration],
  %``Measurement of the cross section for the production of a W boson in association with b-jets in pp collisions at sqrt(s) = 7 TeV with the ATLAS detector,''
  arXiv:1109.1470 [hep-ex].
  %%CITATION = ARXIV:1109.1470;%%
  
  %\cite{Aad:2011jn}
\bibitem{Aad:2011jn}
  G.~Aad {\it et al.}  [ATLAS Collaboration],
  %``Measurement of the cross-section for b-jets produced in association with a Z boson at sqrt(s)=7 TeV with the ATLAS detector,''
  Phys.\  Lett.\  B {\bf 706} (2012) 295
  [arXiv:1109.1403 [hep-ex]].
  %%CITATION = ARXIV:1109.1403;%%

\bibitem{CMS-PAS-EWK-10-015}
  CMS Collaboration,
  %``Observation of Z+b,''
  Physics Analysis Summary CMS-PAS-EWK-10-015.
  %%CITATION = CMS-PAS-EWK-10-015;%%

\bibitem{CMS-PAS-EWK-11-013}
  CMS Collaboration,
  %``Study of associated charm production in W final states at sqrt s = 7 TeV,''
  Physics Analysis Summary CMS-PAS-EWK-11-013.
  %%CITATION = CMS-PAS-EWK-11-013;%%

%% shapes
% CMS-PAS-QCD-10-014
\bibitem{CMS-PAS-QCD-10-014}
  CMS Collaboration,
  %``Jet Transverse Structure and Momentum Distribution in pp Collisions at 7 TeV,''
  Physics Analysis Summary CMS-PAS-QCD-10-014.

\bibitem{Aad:2011kq}
  G.~Aad {\it et al.}  [Atlas Collaboration],
  %``Study of Jet Shapes in Inclusive Jet Production in pp Collisions at sqrt(s) = 7 TeV using the ATLAS Detector,''
  Phys.\ Rev.\ D {\bf 83} (2011) 052003
  [arXiv:1101.0070 [hep-ex]].
  %%CITATION = ARXIV:1101.0070;%%

%% fragmentation references
\bibitem{Aad:2011sc}
  G.~Aad {\it et al.}  [ATLAS Collaboration],
  %``Measurement of the jet fragmentation function and transverse profile in proton-proton collisions at a center-of-mass energy of 7 TeV with the ATLAS detector,''
  Eur.\ Phys.\ J.\ C {\bf 71} (2011) 1795
  [arXiv:1109.5816 [hep-ex]].
  %%CITATION = ARXIV:1109.5816;%%

\bibitem{filtering}
  J.~M.~Butterworth, A.~R.~Davison, M.~Rubin and G.~P.~Salam,
  %``Jet substructure as a new Higgs search channel at the LHC,''
  Phys.\ Rev.\ Lett.\  {\bf 100} (2008) 242001
  [arXiv:0802.2470 [hep-ph]].
  %%CITATION = ARXIV:0802.2470;%%

\bibitem{pruning}
  S.~D.~Ellis, C.~K.~Vermilion and J.~R.~Walsh,
  %``Recombination Algorithms and Jet Substructure: Pruning as a Tool for Heavy Particle Searches,''
  Phys.\ Rev.\ D {\bf 81} (2010) 094023
  [arXiv:0912.0033 [hep-ph]].
  %%CITATION = ARXIV:0912.0033;%%

\bibitem{Altheimer:2012mn}
  A.~Altheimer {\it et al.},
  %``Jet Substructure at the Tevatron and LHC: New results, new tools, new
  %benchmarks,''
  arXiv:1201.0008 [hep-ph].
  %%CITATION = ARXIV:1201.0008;%%
  
\bibitem{ATL-PHYS-PUB-2011-013}
ATLAS Collaboration, public ATLAS note ATL-PHYS-PUB-2011-013, Nov 2011.

%\cite{Aad:2010sp}
\bibitem{Aad:2010sp}
  G.~Aad {\it et al.}  [Atlas Collaboration],
  %``Measurement of the inclusive isolated prompt photon cross section in pp collisions at sqrt(s) = 7 TeV with the ATLAS detector,''
  Phys.\ Rev.\ D {\bf 83} (2011) 052005
  [arXiv:1012.4389 [hep-ex]].
  %%CITATION = ARXIV:1012.4389;%%

%\cite{Aad:2011tw}
\bibitem{Aad:2011tw}
  G.~Aad {\it et al.}  [ATLAS Collaboration],
  %``Measurement of the inclusive isolated prompt photon cross-section in pp collisions at sqrt(s)= 7 TeV using 35 pb-1 of ATLAS data,''
  Phys.\ Lett.\ B {\bf 706} (2011) 150
  [arXiv:1108.0253 [hep-ex]].
  %%CITATION = ARXIV:1108.0253;%%

%\cite{Khachatryan:2010fm}
\bibitem{Khachatryan:2010fm}
  V.~Khachatryan {\it et al.}  [CMS Collaboration],
  %``Measurement of the Isolated Prompt Photon Production Cross Section in $pp$ Collisions at $\sqrt{s} = 7$~TeV,''
  Phys.\ Rev.\ Lett.\  {\bf 106} (2011) 082001
  [arXiv:1012.0799 [hep-ex]].
  %%CITATION = ARXIV:1012.0799;%%

%\cite{Chatrchyan:2011qt}
\bibitem{Chatrchyan:2011qt}
  S.~Chatrchyan {\it et al.}  [CMS Collaboration],
  %``Measurement of the Production Cross Section for Pairs of Isolated Photons in pp collisions at sqrt(s) = 7 TeV,''
  arXiv:1110.6461 [hep-ex].
  %%CITATION = ARXIV:1110.6461;%%  

%\cite{Aad:2011mh}
\bibitem{Aad:2011mh}
  G.~Aad {\it et al.}  [ATLAS Collaboration],
  %``Measurement of the isolated di-photon cross-section in pp collisions at sqrt(s) = 7 TeV with the ATLAS detector,''
  Phys.\ Rev.\ D {\bf 85} (2012) 012003
  [arXiv:1107.0581 [hep-ex]].
  %%CITATION = ARXIV:1107.0581;%%
    
\bibitem{Catani:2011qz}
  S.~Catani, L.~Cieri, D.~de Florian, G.~Ferrera and M.~Grazzini,
  %``Diphoton production at hadron colliders: a fully-differential QCD calculation at NNLO,''
  arXiv:1110.2375 [hep-ph].
  %%CITATION = ARXIV:1110.2375;%%

\bibitem{diphoton-NNLO-dphi}
  S.~Catani, L.~Cieri, D.~de Florian, G.~Ferrera and M.~Grazzini,
  private communication.

%\cite{Butterworth:2010ym}
\bibitem{Butterworth:2010ym}
  A.~Buckley {\it et al.} in 
  ``Les Houches: The Tools and Monte Carlo working group Summary Report,''
  arXiv:1003.1643 [hep-ph] p84-90.
  %%CITATION = ARXIV:1003.1643;%%


\bibitem{fewz}
  R.~Gavin, Y.~Li, F.~Petriello and S.~Quackenbush,
  %``FEWZ 2.0: A code for hadronic Z production at next-to-next-to-leading order,''
  Comput.\ Phys.\ Commun.\  {\bf 182} (2011) 2388
  [arXiv:1011.3540 [hep-ph]].
  %%CITATION = ARXIV:1011.3540;%%

\bibitem{dynnlo}
  S.~Catani, L.~Cieri, G.~Ferrera, D.~de Florian and M.~Grazzini,
  %``Vector boson production at hadron colliders: a fully exclusive QCD
  %calculation at NNLO,''
  Phys.\ Rev.\ Lett.\  {\bf 103} (2009) 082001
  [arXiv:0903.2120 [hep-ph]].
  %%CITATION = PRLTA,103,082001;%%

%\cite{Aad:2011fp}
\bibitem{Aad:2011fp}
  G.~Aad {\it et al.}  [ATLAS Collaboration],
  %``Measurement of the Transverse Momentum Distribution of W Bosons in pp Collisions at sqrt(s) = 7 TeV with the ATLAS Detector,''
  arXiv:1108.6308 [hep-ex].
  %%CITATION = ARXIV:1108.6308;%%  
  
  
%\cite{Aad:2011gj}
\bibitem{Aad:2011gj}
  G.~Aad {\it et al.}  [ATLAS Collaboration],
  %``Measurement of the transverse momentum distribution of Z/gamma* bosons in proton-proton collisions at sqrt(s)=7 TeV with the ATLAS detector,''
  Phys.\ Lett.\ B {\bf 705} (2011) 415
  [arXiv:1107.2381 [hep-ex]].
  %%CITATION = ARXIV:1107.2381;%%

\bibitem{Chatrchyan:2011wt}
  S.~Chatrchyan {\it et al.}  [CMS Collaboration],
  %``Measurement of the Rapidity and Transverse Momentum Distributions of Z Bosons in pp Collisions at sqrt(s)=7 TeV,''
  arXiv:1110.4973 [hep-ex].
  %%CITATION = ARXIV:1110.4973;%%

\bibitem{Aad:2011kt}
  G.~Aad {\it et al.}  [ATLAS Collaboration],
  %``Measurement of the Z to tau tau Cross Section with the ATLAS Detector,''
  Phys.\ Rev.\ D {\bf 84} (2011) 112006
  [arXiv:1108.2016 [hep-ex]].
  %%CITATION = ARXIV:1108.2016;%%

\bibitem{Aad:2011fu}
  G.~Aad {\it et al.}  [ATLAS Collaboration],
  %``Measurement of the W to tau nu Cross Section in pp Collisions at sqrt(s) = 7 TeV with the ATLAS experiment,''
  Phys.\ Lett.\ B {\bf 706} (2012) 276
  [arXiv:1108.4101 [hep-ex]].
  %%CITATION = ARXIV:1108.4101;%%

\bibitem{Chatrchyan:2011nv}
  S.~Chatrchyan {\it et al.}  [CMS Collaboration],
  %``Measurement of the Inclusive Z Cross Section via Decays to Tau Pairs in $pp$ Collisions at $\sqrt{s}=7$ TeV,''
  JHEP {\bf 1108} (2011) 117
  [arXiv:1104.1617 [hep-ex]].
  %%CITATION = ARXIV:1104.1617;%%


\bibitem{Chatrchyan:2011ig}
  S.~Chatrchyan {\it et al.}  [CMS Collaboration],
  %``Measurement of the Polarization of W Bosons with Large Transverse Momenta in W+Jets Events at the LHC,''
  Phys.\ Rev.\ Lett.\  {\bf 107} (2011) 021802
  [arXiv:1104.3829 [hep-ex]].
  %%CITATION = ARXIV:1104.3829;%%


\bibitem{ATLAS-CONF-2011-129}
  ATLAS Collaboration,
  %``An extrapolation to a larger fiducial volume of the measurement of the W-&amp;gt;lnu charge asymmetry in proton-proton collisions at sqrt(s)=7TeV with the ATLAS detector,''
  Conference Note ATLAS-CONF-2011-129.

\bibitem{CMS-Twiki-EWK}
{\tt https://twiki.cern.ch/twiki/bin/view/CMSPublic/PhysicsResultsEWK}

 
 %\cite{Aad:2011dm}
\bibitem{Aad:2011dm}
  G.~Aad {\it et al.}  [ATLAS Collaboration],
  %``Measurement of the inclusive W+- and Z/gamma cross sections in the electron and muon decay channels in pp collisions at sqrt(s) = 7 TeV with the ATLAS detector,''
  arXiv:1109.5141 [hep-ex].
  %%CITATION = ARXIV:1109.5141;%% 

%\cite{Aad:2012qf}
\bibitem{Aad:2012qf}
  G.~Aad {\it et al.}  [ATLAS Collaboration],
  %``Measurement of the top quark pair production cross-section with ATLAS in the single lepton channel,''
  arXiv:1201.1889 [hep-ex].
  %%CITATION = ARXIV:1201.1889;%%
  
 %\cite{Chatrchyan:2011yy}
\bibitem{Chatrchyan:2011yy}
  S.~Chatrchyan {\it et al.}  [CMS Collaboration],
  %``Measurement of the t $\bar{t} Production Cross Section in pp Collisions at 7 TeV in Lepton + Jets Events Using b-quark Jet Identification,''
  Phys.\ Rev.\ D {\bf 84} (2011) 092004
  [arXiv:1108.3773 [hep-ex]].
  %%CITATION = ARXIV:1108.3773;%%

% CMS-PAS-TOP-11-024
\bibitem{CMS-PAS-TOP-11-024}
  CMS Collaboration,
  %``Combination of top pair production cross section measurements,''
  Physics Analysis Summary CMS-PAS-TOP-11-024.

% ATLAS-CONF-2011-108
\bibitem{ATLAS-CONF-2011-108}
  ATLAS Collaboration,
  %``Measurement of the top quark pair production cross-section based on a statistical combination of measurements of dilepton and single-lepton final states at $\sqrt{s} = 7$~TeV with the ATLAS detector,''
  Conference Note ATLAS-CONF-2011-108.
  
\bibitem{CMS-TOP-Twiki}  
  {\tt https://twiki.cern.ch/twiki/bin/view/CMSPublic/PhysicsResultsTOPSummaryPlots}

% ATLAS-CONF-2011-142
\bibitem{ATLAS-CONF-2011-142}
  ATLAS Collaboration,
  %``Reconstructed jet multiplicities from the top-quark pair decays and associated jets in pp collisions at $\sqrt{s} = 7$~TeV measured with the ATLAS detector at the LHC,''
  Conference Note ATLAS-CONF-2011-142.

\bibitem{Plehn:2011tg}
  T.~Plehn and M.~Spannowsky,
  %``Top Tagging,''
  arXiv:1112.4441 [hep-ph].
  %%CITATION = ARXIV:1112.4441;%%

% CMS-PAS-EXO-11-006
\bibitem{CMS-PAS-EXO-11-006}
  CMS Collaboration,
  %``Search for BSM \ttbar Production in the Boosted All-Hadronic Final State,''
  Physics Analysis Summary CMS-PAS-EXO-11-006.

% ATLAS-CONF-2011-153
\bibitem{ATLAS-CONF-2011-153}
  ATLAS Collaboration,
  %``Measurement of the inclusive $t\bar t \gamma$ cross section with the ATLAS detector,''
  Conference Note ATLAS-CONF-2011-153.

% CMS-PAS-TOP-10-009
\bibitem{CMS-PAS-TOP-10-009}
  CMS Collaboration,
  %``Measurement of the top quark mass in the l+jets channel,''
  Physics Analysis Summary CMS-PAS-TOP-10-009.

% ATLAS-CONF-2011-120
\bibitem{ATLAS-CONF-2011-120}
  ATLAS Collaboration,
  %``Measurement of the top quark mass from 2011 ATLAS data using the template method,''
  Conference Note ATLAS-CONF-2011-120.

% CMS-PAS-TOP-11-008
\bibitem{CMS-PAS-TOP-11-008}
  CMS Collaboration,
  %``Determination of the Top Quark Mass from the ttbar Cross Section at sqrt(s) = 7 TeV,''
  Physics Analysis Summary CMS-PAS-TOP-11-008.

\bibitem{ATLAS-CONF-2011-054}
  ATLAS Collaboration,
  %``Determination of the Top-Quark Mass from the $t \bar t$ Cross Section Measurement in pp Collisions at $\sqrt{s}=7$~TeV with the ATLAS detector,''
  Conference Note ATLAS-CONF-2011-054.

% CMS-PAS-TOP-11-019
\bibitem{CMS-PAS-TOP-11-019}
  CMS Collaboration,
  %``Measurement of the mass difference between top and antitop quarks,''
  Physics Analysis Summary CMS-PAS-TOP-11-019.


%\cite{Chatrchyan:2011rr}
\bibitem{Chatrchyan:2011rr}
  S.~Chatrchyan {\it et al.}  [CMS Collaboration],
  %``Measurement of $W\gamma$ and $Z\gamma$ production in $pp$ collisions at $\sqrt{s} = 7$ TeV,''
  Phys.\ Lett.\ B {\bf 701} (2011) 535
  [arXiv:1105.2758 [hep-ex]].
  %%CITATION = ARXIV:1105.2758;%%

%\cite{Aad:2011tc}
\bibitem{Aad:2011tc}
  G.~Aad {\it et al.}  [ATLAS Collaboration],
  %``Measurement of Wgamma and Zgamma production in proton-proton collisions at sqrt(s)=7 TeV with the ATLAS Detector,''
  JHEP {\bf 1109} (2011) 072
  [arXiv:1106.1592 [hep-ex]].
  %%CITATION = ARXIV:1106.1592;%%

%\cite{Chatrchyan:2011tz}
\bibitem{Chatrchyan:2011tz}
  S.~Chatrchyan {\it et al.}  [CMS Collaboration],
  %``Measurement of W+W- Production and Search for the Higgs Boson in pp Collisions at sqrt(s) = 7 TeV,''
  Phys.\ Lett.\ B {\bf 699} (2011) 25
  [arXiv:1102.5429 [hep-ex]].
  %%CITATION = ARXIV:1102.5429;%%

%\cite{Aad:2011kk}
\bibitem{Aad:2011kk}
  G.~Aad {\it et al.}  [ATLAS Collaboration],
  %``Measurement of the WW cross section in sqrt(s) = 7 TeV pp collisions with ATLAS,''
  Phys.\ Rev.\ Lett.\  {\bf 107} (2011) 041802
  [arXiv:1104.5225 [hep-ex]].
  %%CITATION = ARXIV:1104.5225;%%

%\cite{Aad:2011xj}
\bibitem{Aad:2011xj}
  G.~Aad {\it et al.}  [ATLAS Collaboration],
  %``Measurement of the ZZ production cross section and limits on anomalous neutral triple gauge couplings in proton-proton collisions at sqrt(s) = 7 TeV with the ATLAS detector,''
  arXiv:1110.5016 [hep-ex].
  %%CITATION = ARXIV:1110.5016;%%

%\cite{Aad:2011cx}
\bibitem{Aad:2011cx}
  G.~Aad {\it et al.}  [ATLAS Collaboration],
  %``Measurement of the WZ production cross section and limits on anomalous triple gauge couplings in proton-proton collisions at sqrt(s) = 7 TeV with the ATLAS detector,''
  arXiv:1111.5570 [hep-ex].
  %%CITATION = ARXIV:1111.5570;%%


\bibitem{ATLAS-summary}
{\tt https://twiki.cern.ch/twiki/bin/view/AtlasPublic/CombinedSummaryPlots}



\end{thebibliography}
\end{document}